\documentclass[twocolumn,a4paper,reprint,amssymb,aps,prd,groupedaddress,superscriptaddress,nofootinbib]{revtex4-1}
\usepackage[dvips]{graphicx}
\usepackage{amssymb}
\usepackage{filecontents}
\usepackage{amsmath}
\usepackage{color}
\usepackage{float}
\usepackage{textcomp}
\usepackage{bm}
\usepackage{lipsum}

\usepackage{mathtools}

\RequirePackage[colorlinks,citecolor=blue,urlcolor=blue,linkcolor=blue]{hyperref}
\usepackage{yfonts}
\usepackage{amsmath,esint}

\begin{document}
\title {Relativistic binary systems in scale-independent energy-momentum squared gravity} 

\author{\"{O}zg\"{u}r Akarsu}
\email{akarsuo@itu.edu.tr}
\affiliation{Department of Physics, Istanbul Technical University, Maslak 34469 Istanbul, Turkey}

\author{Elham Nazari}
\email{elham.nazari@mail.um.ac.ir}
\affiliation{Department of Physics, Faculty of Science, Ferdowsi University of Mashhad, P.O. Box 1436, Mashhad, Iran}

\author{Mahmood Roshan}
\email{mroshan@um.ac.ir}
\affiliation{Department of Physics, Faculty of Science, Ferdowsi University of Mashhad, P.O. Box 1436, Mashhad, Iran}
\affiliation{School of Astronomy, Institute for Research in Fundamental Sciences (IPM), P.O. Box 19395-5531, Tehran, Iran}

\begin{abstract}
In this paper, we study the gravitational-wave (GW) radiation and radiative behavior of relativistic compact binary systems in the scale-independent energy-momentum squared gravity (EMSG). The field equations of this theory are solved approximately. The gravitational potential of a gravitational source is then obtained by considering two matter Lagrangian densities that both describe a perfect fluid in general relativity (GR). We derive the GW signals emitted from a compact binary system. The results are different from those obtained in general relativity (GR). It is shown that the relevant non-GR corrections modify the wave amplitude and leave the GW polarizations unchanged. 
Interestingly, this modification depends on the choice of the matter Lagrangian density. This means that for different Lagrangian densities, this theory presents different predictions for the GW radiation.
In this case, the system loses energy to modified GWs. This leads to a change in the secular variation of the Keplerian parameters of the binary system. In this work, we investigate the non-GR effects on the radiative parameter, i.e., the first time derivative of the orbital period. Next, applying these results together with GW observations from the relativistic binary systems, we constrain/test the scale-independent EMSG theory in the strong-field regime. 
After assuming that GR is the valid gravity theory, as a priori expectation, we find that the free parameter of the theory is of the order $10^{-5}$ from the direct GW observation, the GW events  GW190425 and GW170817, as well as the indirect GW observation, the double pulsar PSR J0737$-$3039A/B experiment.
\end{abstract}

\maketitle

\section{Introduction}

The standard $\Lambda$ cold dark matter ($\Lambda\text{CDM}$) model has been remarkably successful in explaining the cosmological scale of the Universe~\cite{Riess:1998cb,Planck:2018vyg,Alam:2020sor,DES:2021wwk}. However, the $\Lambda\text{CDM}$ model suffers from a wide range of challenging problems, both theoretically and observationally; it faces the cosmological constant problem~\cite{1989RvMP...61....1W}, coincidence problem~\cite{2014EPJC...74.3160V}, and some persistent tensions, such as $H_0$ and $S_8$ tensions, of various degrees of significance (if not systematics) between some existing data sets from different cosmological and astrophysical probes~\cite{2021APh...13102606D,2021APh...13102605D,2021APh...13102604D,2021APh...13102607D,2022JHEAp..34...49A,2021CQGra..38o3001D,2022NewAR..9501659P,2022PhRvD.106j3506D}. Replacing the general theory of relativity (GR), the standard theory of gravity assumed in the $\Lambda\text{CDM}$ model, is one of the busiest avenues where researchers seek explanations for these challenges facing cosmology today, see, e.g.,~\cite{2010LRR....13....3D,2012PhR...513....1C,2011PhR...509..167C,2016RPPh...79j6901C,2017PhR...692....1N} for reviews on modified theories of gravity.

Recently, a new type of modified theory of gravity has been proposed in which the matter sector of GR is nonlinearly expanded, adding an arbitrary function of the Lorentz scalar $T^{\mu\nu}T_{\mu\nu}$, viz., the self-contraction of the energy-momentum tensor $T^{\mu\nu}$, as $f(T^{\mu\nu}T_{\mu\nu})$ to the Einstein-Hilbert action~\cite{2014EPJP..129..163K,2016PhRvD..94d4002R,2018PhRvD..97b4011A,2017PhRvD..96l3517B}. Referring the mathematical form of the argument of the function $f(T^{\mu\nu}T_{\mu\nu})$, viz., the Lorentz scalar $T^{\mu\nu}T_{\mu\nu}$, this modified theory of gravity has been called energy-momentum squared gravity (EMSG). With the emergence of this modified theory of gravity, it has received a lot of attention and has been studied in different frameworks, see, for instance,~\cite{2018PhRvD..97l4017A,2018PhRvD..98b4031N,2018PhRvD..98f3522A,2020PhRvD.102l4059A,2020PhRvD.102f4016N,2022PhRvD.105d4014N,2020PhRvD.101d4058B,2021EPJP..136..253C,2022PDU....3601013K,2020EPJC...80..150K,2022PhRvD.105j4026N,2022arXiv221004668A,2022PDU....3801128A,2019A&A...625A.127F,2019PhRvD.100h3511B,2020PhRvD.101f4021C,2021PDU....3100774S,2021PDU....3300849R,2022AnPhy.44769149T,2023PDU....4001194A}. One of the specific versions of EMSG is known as energy-momentum powered gravity (EMPG) where $f(T^{\mu\nu}T_{\mu\nu})=\alpha (T^{\mu\nu}T_{\mu\nu})^{\eta}$~\cite{2018PhRvD..97b4011A,2017PhRvD..96l3517B}. Here, both $\alpha$ and $\eta$ are real constants indicating the free parameters of the theory. In this model, $\eta$ determines the power of the EMSG correction and clarifies the energy density scale at which the non-GR contributions can be more effective.  

In the EMPG class, there is a model with $\eta=\frac{1}{2}$ called the scale-independent EMSG~\cite{2018PhRvD..98f3522A}.     
For this particular choice of $\eta$, this theory has the potential to be effective throughout the cosmological epochs with different energy density scales and to play a significant role in early- and late-time evolutions of the Universe. To shed light on this issue, let us choose a perfect fluid to describe the matter sector and examine the Lorentz scalar $(T^{\mu\nu}T_{\mu\nu})^{\eta}$. In this case, the EMPG term proportional to $\rho^{2\eta}$ appears on the right-hand side of the field equations~\cite{2022arXiv221004668A,2018PhRvD..97b4011A}. Here, $\rho$ is the energy density.\footnote {Note that except in Introduction, $\rho$ stands for the rest-mass density in the rest of the paper.}  It is obvious that setting $\eta=\frac{1}{2}$, the matter sector of EMPG and GR will be of the same power, i.e., for both cases, a linear function of the density will appear in the field equations. It does then mean that this correction can play a role along with the GR one at all energy density scales.   
Regarding its interesting and promising applications raised in~\cite{2018PhRvD..98f3522A}, this model deserves to be further investigated in different aspects. In this paper, we focus our attention on the scale-independent EMSG model~\cite{2018PhRvD..98f3522A}.

On the other hand, even the most successful cosmological model needs to survive the gravitational experiments to be considered as a well-founded model. More precisely, their underlying relativistic gravity should pass both weak- and strong-field tests with flying colors. Of course, the EMSG theory and its subclasses are not exempt from this rule. 
The weak-field limit of EMPG and quadratic-EMSG (viz., EMPG with $\eta=1$) is respectively studied in~\cite{2022arXiv221004668A,2022PhRvD.105j4026N}.
It is shown that this theory escapes/passes the solar-system weak-field tests.
In~\cite{2022arXiv221004668A}, it is comprehensively discussed that in this alternative theory, only the mass of an astrophysical object will be modified as $M_{\text{eff}}(\alpha,\,\eta,\,M)=M+M_{\tiny{\text{empg}}}(\alpha,\,\eta,\,M)$ in the relevant weak-field relations where $M$ and $M_{\tiny{\text{empg}}}$ are the physical mass and its EMPG correction, respectively. 
Accordingly, it is expected that the overall behavior of a gravitational system in the weak-field regime of the theory be similar to those in GR. For instance, in~\cite{2022PhRvD.105j4026N}, it is shown that in the quadratic-EMSG gravitational lensing scenario based on the weak-field gravity, the EMSG light curves behave similarly to GR ones. It is also discussed that utilizing astronomical observations such as the Einstein time, the physical mass of a gravitational lens may be overestimated or underestimated depending on the choice of the free parameter of the theory. 
In~\cite{2022arXiv221004668A}, it is clarified that EMPG cannot be distinguished from GR using local gravity observations alone, unless complementary information about the physical mass/density of the gravitational system and the free parameters of the theory is available from other cosmological and astronomical/astrophysical observations or phenomena. In fact, if the mass of the gravitational object is inferred only from the local tests, $M_{\text{eff}}$ and $M$, and consequently, EMPG and GR are not distinguishable. In other words, EMPG escapes weak-field tests. In~\cite{2022PhRvD.105j4026N}, with the implicit assumption that the density of the compact system can be determined from another window, one of us utilizes weak-field experiments to constrain the quadratic-EMSG free parameter.

It is also deserved to ask whether this kind of modified theory can pass the strong-field gravity tests. 
As we know, binary systems which consist of at least one neutron star contain gravitationally strong regions. In light of the accurate measurement of relativistic and radiative effects in them, binary systems provide a suitable testbed for strong-field gravity. In~\cite{2022PhRvD.105d4014N}, utilizing binary pulsars observations, the quadratic-EMSG model is tested in the strong-field regime. It is shown that an EMSG correction depending on the free parameter of the theory as well as the density of the components appears in the first time derivative of the orbital period of the binary system. Using six known binary pulsar experiments and choosing the nuclear density for pulsars, the free parameter of quadratic-EMSG is estimated. In the current paper, in a similar fashion, to probe its viability, we test the scale-independent EMSG model in the strong-field regime of relativistic compact binary systems by studying the gravitational-wave (GW) radiation and the radiative parameter. Given the highly dynamic nature of these systems, it is not obvious how the non-GR corrections of this theory will appear in the relevant strong-field relations and if their role will be similar to those obtained in the weak-field regime. Here, to reveal that, we detail the calculations.

The paper is organized as follows.  The standard formulation of the EMPG field equations is introduced in Sec.~\ref{sec1}. Moreover, in this section, Landau-Lifshitz formalism is employed to reformulate the EMPG field equations. It is a convenient way to derive the post-Minkowskian (PM) and post-Newtonian (PN) expansions of the theory.  As usual, the obtained field equations are highly nonlinear and exceedingly complicated. To solve them approximately, in Sec.~\ref{sec2}, we introduce the PM limit of the scale-independent EMSG model. The gravitational potential in the wave zone of a gravitational source is obtained in this section. Sec.~\ref{sec3} is devoted to the GW radiation in the scale-independent EMSG model. Applying the results obtained in Sec.~\ref{sec3} together with GW observations from the relativistic binary systems, we attempt to constrain/test the asked model in Sec.~\ref{sec4}. Here, we utilize both direct and indirect observations provided by the GW observatories like LIGO and Virgo, and the binary pulsar experiments, respectively.  Our conclusions are presented in Sec.~\ref{Summary}, while Appendix~\ref{app1} summarizes the form of field equation solutions. 

In this paper, $\eta_{\mu\nu}=\text{diag}(-1,\,1,\,1,\,1)$ and Latin and Greek indices run over the values $\{1, 2, 3\}$ and $\{0, 1, 2, 3\}$, respectively.

\section{Scale-independent EMSG} \label{sec1}

In this section, we introduce the EMSG field equations for the specific function $f(T_{\mu\nu}T^{\mu\nu})=\alpha \big(T_{\mu\nu}T^{\mu\nu}\big)^{\eta}$ in the standard 
and Landau-Lifshitz formalisms. This model is called EMPG~\cite{2018PhRvD..97b4011A,2017PhRvD..96l3517B}. Here, $T_{\mu\nu}$ is the energy-momentum tensor and both $\alpha$ and $\eta$ are real constants indicating the free parameters of EMPG modification to GR. It should be mentioned that the dimension of $\alpha$ depends on the value of $\eta$.
We consider that the gravitational system is described by a perfect fluid.
Since the EMPG theory may present different predictions for different matter Lagrangian densities\footnote{Considering Eq. \eqref{teta}, it is seen that the right-hand side of the EMPG field equations \eqref{fieldeq} can be different depending on the choice of the Lagrangian density.}, in this work, we examine two matter Lagrangian densities truly describing a perfect fluid in GR \cite{1970PhRvD...2.2762S,1993CQGra..10.1579B}.

\subsection{Standard formulation}

We first introduce the EMPG field equations in the standard formalism.  For the function $f(T_{\mu\nu}T^{\mu\nu})=\alpha \big(T_{\mu\nu}T^{\mu\nu}\big)^{\eta}$, the Einstein-Hilbert action is modified as
\begin{align}
S=\int \sqrt{-g}\left(\frac{1}{2k}R+\alpha\,\big(T^{\gamma\lambda}T_{\gamma\lambda}\big)^{\eta}+\mathcal{L}_{\text{m}}\right){\rm d}^4x,
\end{align} 
where $g$ is the determinant of the spacetime metric $g_{\mu\nu}$, $k=8\pi G/c^4$ ($G$ being the Newton's constant and $c$ being the speed of light), $R$ is the Ricci scalar, and $\mathcal{L}_{\text{m}}$ is the matter Lagrangian density associated with the energy-momentum tensor $T_{\gamma\lambda}$. Varying this action with respect to the inverse metric $g^{\mu\nu}$, we arrive at 
\begin{align}\label{fieldeq}
G_{\mu\nu}=k T_{\mu\nu}^{\text{eff}},
\end{align}
where $G_{\mu\nu}$ is the Einstein tensor and
\begin{align}\label{T_eff}
T_{\mu\nu}^{\text{eff}}=T_{\mu\nu}+\alpha\,\Big(T^{\gamma\lambda}T_{\gamma\lambda}\Big)^{\eta}\Big[g_{\mu\nu}-2\eta\,\frac{\theta_{\mu\nu}}{T^{\gamma\lambda}T_{\gamma\lambda}}\Big],
\end{align}
is the effective energy-momentum tensor constructed from the standard and the EMPG parts. The standard energy-momentum tensor is defined as 
\begin{align}\label{T_GR}
T_{\mu\nu}=g_{\mu\nu}\mathcal{L}_{\text{m}}-2\frac{\partial\mathcal{L}_{\text{m}}}{\partial g^{\mu\nu}},
\end{align} 
by assuming that $\mathcal{L}_{\text{m}}$ does not depend on metric derivatives, see~\cite{2011PhRvD..84b4020H} and references therein. 
Hereafter, we call the second part of Eq.~\eqref{T_eff} the EMPG energy-momentum tensor
\begin{align}\label{T_EMPG}
T_{\mu\nu}^{\text{\tiny EMPG}}=\alpha\,\Big(T^{\gamma\lambda}T_{\gamma\lambda}\Big)^{\eta}\Big[g_{\mu\nu}-2\eta\,\frac{\theta_{\mu\nu}}{T^{\gamma\lambda}T_{\gamma\lambda}}\Big].
\end{align}
In the above relation, the tensor $\theta_{\mu\nu}$ is defined as  
\begin{align}\label{teta}
\nonumber
\theta_{\mu\nu}=&-2\mathcal{L}_{\text{m}}\Big(T_{\mu\nu}-\frac{1}{2}T\,g_{\mu\nu}\Big)-T\,T_{\mu\nu}+2T^{\gamma}_{\mu}T_{\nu\gamma}\\
&-4T^{\gamma\lambda}\frac{\partial^2\mathcal{L}_{\text{m}}}{\partial g^{\mu\nu}\partial g^{\gamma\lambda}},
\end{align}
where $T=g_{\gamma\lambda}T^{\gamma\lambda}$.

Regarding the Bianchi identities, one can deduce that in this theory, the effective energy-momentum tensor is conserved: 
\begin{align}\label{eq_con}
\nabla_{\mu} T^{\mu\nu}_{\text{eff}}=0.
\end{align}
This relation means that
\begin{align}\label{eq_con_1}
\nonumber
\nabla_{\mu} T^{\mu\nu}=&-\alpha g^{\mu\nu}\nabla_{\mu}\left(T^{\gamma\lambda}T_{\gamma\lambda}\right)^{\eta}\\
&+2\alpha\eta\nabla_{\mu}\left(\frac{\theta^{\mu\nu}}{\left(T^{\gamma\lambda}T_{\gamma\lambda}\right)^{1-\eta}}\right).
\end{align} 
As seen, the standard energy-momentum tensor is not necessarily  conserved in this theory. To specify one of its consequences, let us indicate the matter source.  
We consider that the gravitating system is described by a perfect fluid with 
\begin{align}
T_{\mu\nu}=\left(\varepsilon+\frac{1}{c^2}p\right)u_{\mu}u_{\nu}+p\,g_{\mu\nu},
\end{align}
where $\varepsilon=\rho\big(1+\frac{1}{c^2}\Pi\big)$ is the energy density, $p$ is the pressure, and $u^{\mu}=\gamma(c,\boldsymbol{v})$ is the four-velocity field. Here, $\rho$ is the rest-mass density of a fluid element, $\rho\Pi$ is the proper internal energy density ($\Pi$ then is the internal energy per unit mass), $\gamma=u^0/c$, and $\boldsymbol{v}$ is the three-velocity field.
Furthermore, we consider two Lagrangian densities $\mathcal{L}_{\text{m}}=p$ and $\mathcal{L}_{\text{m}}=-\varepsilon c^2$. These Lagrangian densities correctly describe a perfect fluid in GR~\cite{1970PhRvD...2.2762S,1993CQGra..10.1579B}.
We take the advantage of setting the last term of the tensor $\theta_{\mu\nu}$ to zero, viz., $\partial^2\mathcal{L}_{\text{m}}/\partial g^{\mu\nu}\partial g^{\gamma\lambda}=0$~\cite{2017PhRvD..96l3517B,2011PhRvD..84b4020H,2013PhRvD..88d4023H,2013PhLB..725..437O,Akarsu2023Return}. We also restrict ourselves to the scale-independent EMSG, corresponding to the case
\begin{align}
\eta=\frac{1}{2}
\end{align}
of EMPG, for reasons discussed in the Introduction section. In this case, the coupling parameter $\alpha$ becomes dimensionless. 
At this point, we need to mention another feature of this model, similar to the fact that conservation of the energy-momentum tensor is not necessary in this model. For instance, for the case $\mathcal{L}_{\text{m}}=p$, using Eq.~\eqref{eq_con_1}, one can deduce that 
\begin{align}\label{eq_con_2}
\nabla_{\mu}\left(\varepsilon u^{\mu}\right)=\frac{\alpha}{1+\alpha}u^{\mu}\nabla_{\mu}\varepsilon,
\end{align}
for a dust fluid. It means that unlike GR, it is not necessary that the matter-current conservation is satisfied, i.e.,  $\nabla_{\mu}\left(\varepsilon u^{\mu}\right)$ vanishes, in this model. On the other hand, in this paper, which focuses on an astrophysical investigation of the model, we proceed with the assumption 
\begin{equation} \label{eqn:baryonum}
\nabla_{\mu}\left(\rho u^{\mu}\right)=0,   
\end{equation}
implying baryon number conservation~\cite{poisson2014gravity,will2018theory}. This assumption is fully compatible with Eq.~\eqref{eq_con_2}, although at first glance it may not seem so. The argument on this point is briefly as follows. We can straightforwardly rewrite Eq.~\eqref{eq_con_2} as $
\nabla_{\mu}\left(\varepsilon u^{\mu}\right)=-\alpha \varepsilon \nabla_{\mu}u^{\mu}$, and then, taking $u^{\mu}=c \delta^{\mu}_0$ here, reach $\nabla_{\mu}\left(\varepsilon u^{\mu}\right)=-3 \alpha H\varepsilon$, where $H$ is the Hubble function, in the cosmological context. Accordingly, ignoring the relativistic corrections such as the internal energy, we end up with $\nabla_{\mu}\left(\rho u^{\mu}\right)=-3 \alpha H \rho$. This last equation tells us that in this model, there is matter creation/annihilation in cosmic fluid in an expanding ($H>0$) universe, implying the number of baryons is not conserved on cosmological scales (see~\cite{2023PDU....4001194A} for cosmological consequences of this feature of the scale-independent EMSG). This seems to be incompatible with Eq.~\eqref{eqn:baryonum}. However, the galaxies (each is a gravitationally bound system consisting of stars, cold dark matter, etc.) themselves are independent of the expansion of the universe, meaning they do not expand (i.e., $H=0$ in the local region of the universe occupied by a galaxy), and thus matter creation/annihilation does not occur within galaxies, but in the expanding space between the galaxies. Consequently, well inside the galaxies, and therefore also for astrophysical objects located in a galaxy, such as the binary stars we are dealing with in this paper, the conservation of the baryon number, Eq.~\eqref{eqn:baryonum}, would exactly apply. This also allows us to write
\begin{align}\label{rhos}
\partial_t\rho^*+\partial_j\big(\rho^* v^{j}\big)=0,
\end{align}
where $\rho^*=\sqrt{-g}\gamma \rho$, which will help simplify our calculations later in the paper.

\subsection{Landau-Lifshitz formulation}

In order to study radiative aspects of EMPG, similar to~\cite{2022PhRvD.105d4014N}, we utilize the Landau-Lifshitz formalism. It can be shown that the Landau-Lifshitz formulation of the EMPG field equations~\eqref{fieldeq} is given by 
\begin{align}\label{fieldeq_LL}
\square h^{\mu\nu}=-2\,k\,\tau^{\mu\nu}_{\text{eff}}.
\end{align}
Here, the harmonic gauge conditions 
\begin{align}\label{harmonic-gauge}
\partial_{\mu}h^{\mu\nu}=0,
\end{align}
are imposed.
In this approach, $h^{\mu\nu}$, which is a function of harmonic (or de Donder) coordinates,\footnote{In another coordinate system, the wave equations should be corrected. For instance, for the case of radiative coordinates, see  \cite{2023CQGra..40e5006T}.} is the gravitational potential representing the deviation of the gothic metric, $\mathfrak{g}^{\mu\nu}=\sqrt{-g}g^{\mu\nu}$, from the Minkowski metric $\eta^{\mu\nu}$, i.e., $\mathfrak{g}^{\mu\nu}=\eta^{\mu\nu}-h^{\mu\nu}$. In Eq.~\eqref{fieldeq_LL},  $\square=\eta^{\gamma\lambda}\partial_{\gamma\lambda}$ and $\tau^{\mu\nu}_{\text{eff}}$ is the effective energy-momentum pseudotensor which is written as 
\begin{align}\label{tau}
\tau^{\mu\nu}_{\text{eff}}=\big(-g\big)\Big(T^{\mu\nu}_{\text{eff}}+t^{\mu\nu}_{\text{LL}}+t^{\mu\nu}_{\text{H}}\Big).
\end{align}
This pseudotensor is built from Eq.~\eqref{T_eff}, the harmonic pseudotensor
\begin{align}\label{t_H}
(-g)t_{\text{H}}^{\mu\nu}=\frac{1}{2\,k}\big(\partial_\gamma h^{\mu\lambda}\partial_\lambda h^{\nu\gamma}-h^{\gamma\lambda}\partial_{\gamma\lambda}h^{\mu\nu}\big).
\end{align}
and the Landau-Lifshitz pseudotensor whose general definition in terms of the gothic metric $\mathfrak{g}^{\alpha\beta}$ is given by
\begin{align}\label{tLLg}
\nonumber
&(-g)t_{\text{LL}}^{\alpha\beta}=\frac{1}{2\,k}\bigg\lbrace\partial_{\lambda}\mathfrak{g}^{\alpha\beta}\partial_{\mu}\mathfrak{g}^{\lambda\mu}-\partial_{\lambda}\mathfrak{g}^{\alpha\lambda}\partial_{\mu}\mathfrak{g}^{\beta\mu}+\frac{1}{2}g^{\alpha\beta}g_{\lambda\mu}\partial_{\rho}\\\nonumber
&\times\mathfrak{g}^{\lambda\nu}\partial_{\nu}\mathfrak{g}^{\mu\rho}-g^{\alpha\lambda}g_{\mu\nu}\partial_{\rho}\mathfrak{g}^{\beta\nu}\partial_{\lambda}\mathfrak{g}^{\mu\rho}-g^{\beta\lambda}g_{\mu\nu}\partial_{\rho}\mathfrak{g}^{\alpha\nu}\partial_{\lambda}\mathfrak{g}^{\mu\rho}\\\nonumber
&+g_{\lambda\mu}g^{\nu\rho}\partial_{\nu}\mathfrak{g}^{\alpha\lambda}\partial_{\rho}\mathfrak{g}^{\beta\mu}+\frac{1}{8}\big(2g^{\alpha\lambda}g^{\beta\mu}-g^{\alpha\beta}g^{\lambda\mu}\big)\big(2g_{\nu\rho}g_{\sigma\tau}\\
&-g_{\rho\sigma}g_{\nu\tau}\big)\partial_{\lambda}\mathfrak{g}^{\nu\tau}\partial_{\mu}\mathfrak{g}^{\rho\sigma}\bigg\rbrace.
\end{align}
It reduces to
\begin{align}\label{tLL}
\nonumber
&(-g)t_{\text{LL}}^{\mu\nu}=\frac{1}{2\,k}\bigg\lbrace\frac{1}{2}\eta^{\mu\nu}\eta_{\lambda\gamma}\partial_{\rho}h^{\lambda\sigma}\partial_{\sigma}h^{\gamma\rho}-\eta^{\mu\lambda}\eta_{\gamma\sigma}\partial_{\rho}h^{\nu\sigma}\\\nonumber
&\times\partial_{\lambda}h^{\gamma\rho}-\eta^{\nu\lambda}\eta_{\gamma\sigma}\partial_{\rho}h^{\mu\sigma}\partial_{\lambda}h^{\gamma\rho}+\eta_{\lambda\gamma}\eta^{\sigma\rho}\partial_{\sigma}h^{\mu\lambda}\partial_{\rho}h^{\nu\gamma}\\
&+\frac{1}{8}\big(2\eta^{\mu\lambda}\eta^{\nu\gamma}-\eta^{\mu\nu}\eta^{\lambda\gamma}\big)\big(2\eta_{\epsilon\rho}\eta_{\sigma\tau}-\eta_{\rho\sigma}\eta_{\epsilon\tau}\big)\partial_{\lambda}h^{\epsilon\tau}\partial_{\gamma}h^{\rho\sigma}\bigg\rbrace,
\end{align}
after using the harmonic gauge condition and truncating the results to the leading order $h^{\mu\nu}$. This relation is sufficient for the following calculations in the second PM approximation. 
It should be noted that imposing the condition \eqref{harmonic-gauge} is equivalent to applying the conservation equation
\begin{align}\label{conservation-equation}
\partial_\mu\tau^{\mu\nu}_{\text{eff}}=0.
\end{align}
The way to derive Eqs.~\eqref{fieldeq_LL}-\eqref{tLL} has been discussed in GR and EMSG by~\cite{poisson2014gravity} and~\cite{2022PhRvD.105d4014N}, respectively. We shall not repeat it here and refer the interested reader to these references for the underlying details.

\subsection{Matter source}

To define the EMPG correction term, we specify the matter source. As mentioned earlier, we consider that the gravitating system is described by a perfect fluid.
Also, we assume that this fluid is a PN system where the following conditions   
\begin{align}\label{condi1}
\frac{p}{\rho c^2}\sim\frac{\Pi}{c^2}\sim\frac{v^2}{c^2}\sim\frac{U}{c^2}\ll 1,
\end{align}
are satisfied; namely, the system under consideration is subjected to the slow-motion and weak-field conditions. Here, $U$ is the Newtonian potential. The order of smallness of these four dimensionless quantities is denoted by $O(c^{-2})$. In this work, we focus our attention on the compact-support source. We consider compact bodies which have negligible multipole moments. In the following, we will then examine binary systems of compact objects.

To estimate the PN order of each components of the effective energy-momentum tensor, we study the matter distribution in the flat spacetime, i.e., we have $T_{\mu\nu}=\big(\varepsilon+\frac{1}{c^2}p\big)u_{\mu}u_{\nu}+p\,\eta_{\mu\nu}$. Regarding the conditions listed in Eq.~\eqref{condi1}, one can show that the components of the standard energy-momentum tensor up to the leading PN order are given by
\begin{subequations}
\begin{align}
\label{T00}
&T_{00}=\rho\,c^2,\\
&T_{0j}=-\rho\,c\,v_j,\\
\label{Tij}
&T_{ij}=\rho\,v_iv_j+p\,\delta_{ij}.
\end{align}
\end{subequations}
To examine the EMPG part of the effective energy-momentum tensor, we should specify the Lagrangian density. As mentioned previously, we take both cases $\mathcal{L}_{\text{m}}=-\varepsilon c^2$ and $\mathcal{L}_{\text{m}}=p$ to describe a perfect fluid. To indicate each case, hereafter, we add the indexes ``$\varepsilon$'' and ``$p$'' to the relevant quantities for the cases $\mathcal{L}_{\text{m}}=-\varepsilon c^2$ and $\mathcal{L}_{\text{m}}=p$, respectively.

For $\mathcal{L}_{\text{m}}=-\varepsilon c^2$, by inserting the energy-momentum tensor of the perfect fluid in Eq.~\eqref{T_EMPG}, we find that the time-time component of $T^{\text{\tiny EMPG}}_{\mu\nu}$ is
\begin{align}
\nonumber
&\prescript{\varepsilon}{} T^{\text{\tiny EMPG}}_{00}=-\alpha\,\rho^{2\eta}c^{4\eta}\bigg(\Big(1+\frac{\Pi}{c^2}\Big)^2+\frac{3\,p^2}{\rho^2c^4}\bigg)^{\eta-1}\bigg(\Big(1+4\,\eta\,\frac{v^2}{c^2}\Big)\\
&\times\Big(1+\frac{\Pi}{c^2}\Big)^2+\frac{p}{\rho\,c^2}\Big(\frac{3p}{\rho\,c^2}+8\eta\,\frac{v^2}{c^2}+4\eta\,\frac{v^2}{c^2}\frac{p}{\rho\,c^2}+8\eta\,\frac{v^2}{c^2}\frac{\Pi}{c^2}\Big)\bigg).
\end{align}
To simplify the above relation, the normalization condition $u_{\gamma}u^{\gamma}=-c^2$ is utilized.
After imposing the PN conditions~\eqref{condi1}, it can be shown that this component as well as the rest are reduced as follows:   
\begin{subequations}
\begin{align}
\label{T00_rho}
&\prescript{\varepsilon}{}T^{\text{\tiny EMPG}}_{00}=-\alpha\rho^{2\eta}c^{4\eta}\Big(1+O(c^{-2})\Big),\\
&\prescript{\varepsilon}{}T^{\text{\tiny EMPG}}_{0j}=2\,\alpha\,\eta\,\frac{v_j}{c}\rho^{2\eta}c^{4\eta}\Big(1+O(c^{-2})\Big),\\
\label{Tij_rho}
&\prescript{\varepsilon}{}T^{\text{\tiny EMPG}}_{ij}=\alpha(1-2\eta)\delta_{ij}\rho^{2\eta}c^{4\eta}\Big(1+O(c^{-2})\Big),
\end{align}
\end{subequations}
to the leading PN order.

In a similar fashion, in the case $\mathcal{L}_{\text{m}}=p$, we obtain that
\begin{subequations}
\begin{align}
\label{T00_p}
&\prescript{p}{}T^{\text{\tiny EMPG}}_{00}=\alpha(2\eta-1)\rho^{2\eta}c^{4\eta}\Big(1+O(c^{-2})\Big),\\
&\prescript{p}{}T^{\text{\tiny EMPG}}_{0j}=-2\alpha\, \eta\, \frac{v_j}{c}\rho^{2\eta}c^{4\eta}\Big(1+O(c^{-2})\Big),\\
\label{Tij_p}
&\prescript{p}{}T^{\text{\tiny EMPG}}_{ij}=\alpha\delta_{ij}\rho^{2\eta}c^{4\eta}\Big(1+O(c^{-2})\Big).
\end{align}
\end{subequations}
Interestingly, comparison between Eqs.~\eqref{T00_rho}-\eqref{Tij_rho} and Eqs.~\eqref{T00_p}-\eqref{Tij_p} reveals that the EMPG portion of the effective energy-momentum tensor is different for these two Lagrangian densities. Although both of these Lagrangians describe the same system, i.e., a perfect fluid, the EMPG theory can then present different predictions for different matter Lagrangian densities. In~\cite{2009PhRvD..80l4040F,2008PhRvD..78f4036B}, this fact is studied in $f(R)$ theories of gravity. In this work, we attempt to study GWs for both cases $\mathcal{L}_{\text{m}}=- \varepsilon c^2$ and $\mathcal{L}_{\text{m}}=p$ to clarify this point in the EMPG theory as well.

\section{Gravitational potential}\label{sec2}

In order to study GWs, one should have enough information about the gravitational potential in the wave zone of a gravitational source. In the context of the PN gravity, inside a three-dimensional sphere with a radius of the order of the characteristic wavelength of GWs emitted by the source, is called the near zone. Outside this region where the radiation effects are important is the wave zone. 

To find this potential, we approximately solve the EMPG field equations~\eqref{fieldeq_LL} by using the iteration method introduced in~\cite{poisson2014gravity}. Regarding the position of the field and source points, in the Landau-Lifshitz reformulation of GR, the form of the solutions to the field equations is introduced in this reference. As the mathematical form of Eq.~\eqref{fieldeq_LL} is similar to that in GR, we can use these solutions here.
For the sake of convenience, these required solutions are displayed in Appendix~\ref{app1}. 
In~\cite{2022PhRvD.105d4014N} this method is also applied to solve the quadratic-EMSG (viz., EMPG with $\eta=1$) field equations. Although the method applied here is similar to our previous work, it is constructive to mention the calculation path in detail to clarify when and where non-GR corrections may affect the results in EMPG theory. 

Before we get our hands dirty with the iteration method, let us introduce the PM expansion of the metric in terms of the gravitational potentials $h^{\mu\nu}$ which will be needed in the following derivation. To write this expansion up to the sufficient PN order, we first specify the leading order of the energy-momentum tensor components. Henceforth, we set $\eta=\frac{1}{2}$. Given Eqs.~\eqref{T00}-\eqref{Tij} and~\eqref{T00_rho}-\eqref{Tij_rho}, the leading order of $ \prescript{\varepsilon}{}T^{00}_{\text{eff}}$, $ \prescript{\varepsilon}{}T^{0j}_{\text{eff}}$, and $ \prescript{\varepsilon}{}T^{ij}_{\text{eff}}$ is $O(c^2)$, $O(c)$, $O(1)$, respectively. Therefore, for the case $\mathcal{L}_{\text{m}}=-\varepsilon c^2$, $\prescript{\varepsilon}{}h^{00}$, $\prescript{\varepsilon}{}h^{0j}$, and $\prescript{\varepsilon}{}h^{ij}$ are of the order $c^{-2}$, $c^{-3}$, and $c^{-4}$, respectively.
Now, considering the general form of the PM expansion of the metric 
\begin{align}
\nonumber
 g_{\alpha\beta}=&\eta_{\alpha\beta}+h_{\alpha\beta}-\frac{1}{2}h\eta_{\alpha\beta}+h_{\alpha\mu}h^{\mu}_{\beta}-\frac{1}{2}h h_{\alpha\beta}\\\label{PM_metric}
&+\left(\frac{1}{8}h^2-\frac{1}{4}h^{\mu\nu}h_{\mu\nu}\right)\eta_{\alpha\beta}+O(G^3),
\end{align}
and its determinant~\cite{poisson2014gravity}
\begin{align}
\label{PM_g}
& (-g)=1-h+\frac{1}{2}h^2-\frac{1}{2}h^{\mu\nu}h_{\mu\nu}+O(G^3),
\end{align}
as well as the PN order of the gravitational potential components mentioned above, we arrive at 
\begin{subequations}
\begin{align}
\label{g00rho}
&
\prescript{\varepsilon}{}g_{00}=-1+\frac{1}{2}h^{00}-\frac{3}{8}\big(h^{00}\big)^2+\frac{1}{2}h^{kk}+O(c^{-6}),\\
&\prescript{\varepsilon}{}g_{0j}=-h^{0j}+O(c^{-5}),\\
\label{gjkrho}
&\prescript{\varepsilon}{}g_{ij}=\delta_{ij}\Big(1+\frac{1}{2}h^{00}\Big)+O(c^{-4}),\\
\label{detgrho}
&(-\prescript{\varepsilon}{}g)=1+h^{00}+O(c^{-4}),
\end{align}
\end{subequations}
for the case $\mathcal{L}_{\text{m}}=-\varepsilon c^2$.
Here, $h=\eta_{\alpha\beta}h^{\alpha\beta}=-h^{00}+h^{kk}$.
These components are accurate enough to describe a gravitational system with the first PN (1\tiny PN \normalsize) corrections.

Considering Eqs.~\eqref{T00}-\eqref{Tij} and~\eqref{T00_p}-\eqref{Tij_p} for $\eta=\frac{1}{2}$, one can deduce that $ \prescript{p}{}T^{00}_{\text{eff}}$, $ \prescript{p}{}T^{0j}_{\text{eff}}$, and $ \prescript{p}{}T^{ij}_{\text{eff}}$ are respectively of the order $c^{2}$, $c$, and $c^2$. Then, in the case $\mathcal{L}_{\text{m}}=p$, $\prescript{p}{}h^{00}$, $\prescript{p}{}h^{0j}$, and $\prescript{p}{}h^{ij}$ are of the order $c^{-2}$, $c^{-3}$, and $c^{-2}$, respectively. As seen in this case, the space-space component of the gravitational potential is bigger than that in the previous one, i.e., $\prescript{p}{}h^{ij}/\prescript{\varepsilon}{}h^{ij}=O(c^{2})$. Regarding this point and utilizing Eq.~\eqref{PM_metric}, we find that
\begin{subequations}
\begin{align}
\nonumber
&
\prescript{p}{} g_{00}=-1+\frac{1}{2}h^{00}-\frac{3}{8}\big(h^{00}\big)^2+\frac{1}{2}h^{kk}\big(1-\frac{1}{2}h^{00}\big)\\\label{g00p}
&-\frac{1}{8}\big(h^{kk}\big)^2+O(c^{-6}),\\
\label{g0jp}
&\prescript{p}{} g_{0j}=-h^{0j}+O(c^{-5}),\\
\label{gjkp}
&\prescript{p}{} g_{ij}=\delta_{ij}\Big(1+\frac{1}{2}h^{00}\Big)-\frac{1}{2}\delta_{ij}h^{kk}+h^{ij}+O(c^{-4}),\\
\label{gp}
&(-\prescript{p}{}g)=1+h^{00}-h^{kk}+O(c^{-4}),
\end{align}
\end{subequations}
by which one can correctly describe a system in 1\tiny PN \normalsize limit of EMPG for the case $\mathcal{L}_{\text{m}}=p$ and $\eta=\frac{1}{2}$.

In the next parts, we apply the iteration method.
We solve the field equation $\square h^{\mu\nu}_{\text{\tiny(n)}}=-2\,k\,\tau^{\mu\nu}_{\text{eff}\text{\tiny(n-1)}}$ in each iterated step. Here, the index ``$\text{\tiny(n)}$'' refers to the $n^{\rm th}$ iteration. 
The wave equation, independent of the harmonic gauge condition, is known as the relaxed Einstein field equations.
In this method, the source term of the wave equation, i.e., $\tau^{\mu\nu}_{\text{eff}\text{\tiny(n-1)}}$, is obtained in the previous iterated step. So, knowing the source term, the field equation is no longer non-linear and, in principle, it can be integrated straightforwardly. Depending on the degree of accuracy required, the field equation should be solved up to the sufficient iteration.
It is shown that to find the gravitational potential to the leading PN order, we should carry out the calculation to the second iteration. In fact, $h^{\mu\nu}_{\text{\tiny (2)}}$ should be obtained.
In this step as the last step, we are at liberty to impose the harmonic gauge condition $\partial_\mu h_{\text{\tiny(2)}}^{\mu\nu}=0$ or equivalently the conservation equation $\partial_\mu \tau_{\text{eff}\text{\tiny(1)}}^{\mu\nu}=0$ on the iterated solution to the relaxed field equations \cite{poisson2014gravity,will2018theory}.
Here, similar to Poisson and Will's terminology in the standard textbook~\cite{poisson2014gravity}, the gravitational potential whose source point is located in the near (wave) zone is called the near-zone (wave-zone) potential and shown by $h^{\mu\nu}_{\mathcal{N}}$ ($h^{\mu\nu}_{\mathcal{W}}$).

\subsection{Case $\mathcal{L}_{\text{m}}=p$}

In this section, for the case $\mathcal{L}_{\text{m}}=p$, we find $h^{\mu\nu}_{\text{\tiny (2)}}$ in the wave zone. To do so, we attempt to find the correct source term of this potential  in the following parts.

\subsubsection{First iteration}
As the first step in this method, it is assumed that $h^{\mu\nu}_{\text{\tiny (0)}}=0$ and $g^{\mu\nu}_{\text{\tiny (0)}}=\eta^{\mu\nu}$. By using the normalization condition $u_{\gamma}u^{\gamma}=-c^2$, we obtain that $\gamma_{\text{\tiny(0)}}=1+\frac{1}{2}\frac{v^2}{c^2}+O(c^{-4})$. Therefore, we have  $\rho=\big(1-\frac{1}{2}\frac{v^2}{c^2}+O(c^{-4})\big)\rho^*$. Regarding this relation, one can show that the components of Eq.~\eqref{T_eff} are reduced to
 \begin{subequations}
\begin{align}
\label{T000}
&    \prescript{p}{}T^{00}_{\text{eff}\text{\tiny(0)}}=\rho^*c^2+O(1),\\
&    \prescript{p}{}T^{0j}_{\text{eff}\text{\tiny(0)}}=(1+\alpha)\rho^*c\,v^j+O(c^{-1}),\\
\label{Tij0}
&   \prescript{p}{}T^{ij}_{\text{eff}\text{\tiny(0)}}=\alpha \rho^* c^2\,\delta^{ij}+O(1),
\end{align} 
 \end{subequations}
for a perfect fluid in the flat spacetime. Since $h^{\mu\nu}_{\text{\tiny (0)}}=0$, the landau-Lifshitz and harmonic pseudotensors vanish here, see Eqs.~\eqref{t_H} and~\eqref{tLL}. Now, in the first iterated step, one can solve the wave equation $\square h^{\mu\nu}_{\text{\tiny(1)}}=-2\,k\,\tau^{\mu\nu}_{\text{eff}\text{\tiny(0)}}$ to find $h^{\mu\nu}_{\text{\tiny(1)}}$. It should be noted that according to the position of the source point, the gravitational potential consists of two pieces; the near-zone portion ($h^{\mu\nu}_{\mathcal{N}}$) and the wave-zone portion ($h^{\mu\nu}_{\mathcal{W}}$).

We first construct the near-zone potential in the near zone.
In this case, the source and field points both are in the near zone. The near-zone solution of the wave equation is given by~\eqref{hNear}. Using Eqs.~\eqref{T000}-\eqref{Tij0} in this solution, we find that 
\begin{subequations}
\begin{align}
\label{h001}
& \prescript{p}{}h^{00}_{\mathcal{N}\text{\tiny(1)}}=\frac{4}{c^2}U+O(c^{-4}),\\
& \prescript{p}{}h^{0j}_{\mathcal{N}\text{\tiny(1)}}=\frac{4}{c^3}(1+\alpha)U^j+O(c^{-4}),\\
\label{hij1}
& \prescript{p}{}h^{ij}_{\mathcal{N}\text{\tiny(1)}}=\frac{4\alpha}{c^2}U\delta^{ij}+O(c^{-4}),
\end{align} 
 \end{subequations}
where
\begin{subequations}
\begin{align}
\label{U}
& U=G\int_{\mathfrak{M}}\frac{{\rho^*}'}{\rvert{\boldsymbol{x}-\boldsymbol{x}'}\rvert}{\rm d}^3x',\\
& U^j=G\int_{\mathfrak{M}}\frac{{\rho^*}'v'^j}{\rvert{\boldsymbol{x}-\boldsymbol{x}'}\rvert}{\rm d}^3x'.
\end{align}
\end{subequations}
It should be mentioned that in the time-time and space-space components, the term ${\rm d}/{\rm d} t \int_{\mathfrak{M}} \rho^* {\rm d}^3x$ appears in the order $c^{-3}$. Using Eq.~\eqref{rhos}, one can show that this term turns into the surface integral $\oint_{\partial\mathfrak{M}}\rho^* v^j {\rm d}S_j$. On the other hand, the slow-motion assumption dictates that the matter part of the gravitational system should be situated deep within the near zone. As a result, the matter part of the system has no portion on the surface $\partial\mathfrak{M}$ which is the boundary of the near and wave zones and the surface integrals like the one we encounter here would vanish.  

To complete $h^{\mu\nu}_{\text{\tiny(1)}}=h^{\mu\nu}_{\mathcal{N}\text{\tiny(1)}}+h^{\mu\nu}_{\mathcal{W}\text{\tiny(1)}}$, its wave-zone part should also be found. As the matter part of our system does not exist beyond the near zone and $t^{\mu\nu}_{\text{ LL} \text{\tiny(0)}}=0=t^{\mu\nu}_{\text{H}\text{\tiny(0)}}$, one can straightforwardly conclude that $h^{\mu\nu}_{\mathcal{W}\text{\tiny(1)}}=0$ in this step. Therefore, we have  $h^{\mu\nu}_{\text{\tiny(1)}}=h^{\mu\nu}_{\mathcal{N}\text{\tiny(1)}}$.

Given Eqs.~\eqref{h001}-\eqref{hij1}, we can now build the metric components in this step. Substituting these relations back into Eqs.~\eqref{g00p}-\eqref{gp}, we arrive at
  \begin{subequations}
 \begin{align}
 \label{g00p1}
 &  \prescript{p}{}g_{00}^{\text{\tiny(1)}}=-1+\frac{2}{c^2}(1+3\alpha)U+O(c^{-4}),\\
 & \prescript{p}{}g_{0j}^{\text{\tiny(1)}}=-\frac{4}{c^3}(1+\alpha)U^j+O(c^{-4}),\\
 \label{gijp1}
 & \prescript{p}{}g_{ij}^{\text{\tiny(1)}}=\Big(1+\frac{2}{c^2}(1-\alpha)U\Big)\delta^{ij}+O(c^{-4}),\\
 &{(-\prescript{p}{}g^{\text{\tiny(1)}})}=1+\frac{4}{c^2}(1-3\alpha)U+O(c^{-4}).
 \end{align} 
  \end{subequations}
These are all materials, with sufficient PN orders, that we will need in the following step.

\subsubsection{Second iteration}

Keeping in mind the normalization condition and using the components of the metric $\prescript{p}{}g_{\mu\nu}^{\text{\tiny(1)}}$ obtained before, we get
\begin{align}
\prescript{p}{}\gamma_{\text{\tiny (1)}}=1+\frac{1}{c^2}(1+3\alpha)U+\frac{1}{2}\frac{v^2}{c^2}+O(c^{-4}),
\end{align}
and
\begin{align}
\rho^*=\Big[1+\frac{3}{c^2}(1-\alpha)U+\frac{1}{2}\frac{v^2}{c^2}+O(c^{-4})\Big]\rho.
\end{align}
Using these definitions along with Eqs.~\eqref{g00p1}-\eqref{gijp1}, one can obtain the components of the standard and EMPG energy-momentum tensors as  
 \begin{subequations}
\begin{align}
\label{T001}
&\prescript{p}{} T^{00}_{\text{\tiny(1)}}=\rho^*c^2+O(1),\\
& \prescript{p}{}T^{0j}_{\text{\tiny(1)}}=\rho^*c\,v^j+O(c^{-1}),\\
& \prescript{p}{}T^{ij}_{\text{\tiny(1)}}=\rho^{*}v^iv^j+p\,\delta^{ij}+O(c^{-2}),
\end{align} 
 \end{subequations}
and 
 \begin{subequations}
\begin{align}
&\prescript{p}{} T^{00 \text{\tiny(1)}}_{\text{\tiny EMPG}}=O(1),\\
&\prescript{p}{} T^{0j \text{\tiny(1)}}_{\text{\tiny EMPG}}=\alpha \rho^*c\,v^j+O(c^{-1}),\\
\nonumber
& \prescript{p}{}T^{ij \text{\tiny(1)}}_{\text{\tiny EMPG}}=\alpha\rho^*c^2\Big[\Big(1-\frac{1}{c^2}\big(\frac{1}{2}v^2+5(1-\alpha)U-\Pi\big)\Big)\delta^{ij}\\
&+\frac{1}{c^2}v^iv^j\Big]+O(c^{-2}),
\end{align} 
 \end{subequations}
respectively. Here, we truncate the results to the required PN order for the next calculations. Interestingly, a term with an unusual order $O(c^2)$ appears in $\prescript{p}{}T^{ij \text{\tiny(1)}}_{\text{\tiny EMPG}}$. In the following, we trace the possible role of this high-order expression in the gravitational potential $\prescript{p}{}h^{\mu\nu}_{\text{\tiny(2)}}$ and thus in GWs. 
To complete the source term of the wave equation $\square h^{\mu\nu}_{\text{\tiny(2)}}=-2\,k\,\tau^{\mu\nu}_{\text{eff}\text{\tiny(1)}}$ in this step, $(-g)t^{\mu\nu}_{\text{ LL}}$ and $(-g)t^{\mu\nu}_{\text{H}}$ should also be derived. To do so, we insert Eqs.~\eqref{h001}-\eqref{hij1} into the definitions of these pseudotensors. After some manipulations, we finally obtain  
 \begin{subequations}
\begin{align}
&(- \prescript{p}{}g_{\text{\tiny(1)}}) \prescript{p}{}t^{00 \text{\tiny(1)}}_{\text{ LL}}=O(1),\\
& (-\prescript{p}{}g_{\text{\tiny(1)}})\prescript{p}{}t^{0j \text{\tiny(1)}}_{\text{ LL}}=O(c^{-1}),\\
\nonumber
& (-\prescript{p}{}g_{\text{\tiny(1)}})\prescript{p}{}t^{ij \text{\tiny(1)}}_{\text{ LL}}=\frac{1}{4\pi G}\big(1+6\alpha-7\alpha^2\big)\Big(\partial^iU\partial^jU-\frac{1}{2}\delta^{ij}\\
&\times\partial^nU\partial_nU\Big)-\frac{\alpha^2}{\pi\,G}\Big(\partial^iU\partial^jU-\delta^{ij}\partial^nU\partial_nU\Big)+O(c^{-2}),
\end{align} 
 \end{subequations} 
for the Landau-Lifshitz pseudotensor components and 
 \begin{subequations}
\begin{align}
&  (-\prescript{p}{}g_{\text{\tiny(1)}})\prescript{p}{}t^{00 \text{\tiny(1)}}_{\text{ H}}=O(1),\\
&(- \prescript{p}{}g_{\text{\tiny(1)}}) \prescript{p}{}t^{0j \text{\tiny(1)}}_{\text{ H}}=O(c^{-1}),\\
\label{thij1}
& (-\prescript{p}{}g_{\text{\tiny(1)}})\prescript{p}{}t^{ij \text{\tiny(1)}}_{\text{H}}=\frac{\alpha^2}{\pi G}\Big(\partial^iU\partial^jU-\delta^{ij}U \partial_n\partial^nU\Big)+O(c^{-2}),
\end{align} 
 \end{subequations} 
for the components of the harmonic pseudotensor in this iterated step. As seen, at $O(1)$, the space-space component of the harmonic pseudotensor is only made up of the EMPG terms. Gathering together Eqs.~\eqref{T001}-\eqref{thij1} reveals that 
 \begin{subequations}
\begin{align}
\label{tau00p1}
&\prescript{p}{}\tau^{00 \text{\tiny(1)}}_{\text{ eff}}=\rho^*c^2+O(1),\\\label{tau0jp1}
&\prescript{p}{}\tau^{0j \text{\tiny(1)}}_{\text{ eff}}=(1+\alpha)\rho^*c\,v^j+O(c^{-1}),\\
\nonumber
& \prescript{p}{}\tau^{ij \text{\tiny(1)}}_{\text{eff}}=(1+\alpha)\rho^*v^iv^j+\frac{1}{4\pi G}\big(1+6\alpha-7\alpha^2\big)\partial^iU\partial^jU\\\nonumber
&+\delta^{ij}\Big[p+\alpha\rho^*c^2\Big(1-\frac{1}{c^2}\big(\frac{1}{2}v^2+(1+3\alpha)U-\Pi\big)\Big)\\\label{tauijp1}
&-\frac{1}{8\pi G}\big(1+6\alpha-15\alpha^2\big)\partial^nU\partial_nU\Big]+O(c^{-2}).
\end{align} 
 \end{subequations} 
Up to this point, no assumption has been made about the order of magnitude of the theory parameter, $\alpha$.

Knowing that the calculation is carried out to the second iteration, we can now impose $\partial_\mu \tau_{\text{eff}}^{\mu\nu\text{\tiny(1)}}=0$ which is formally equivalent to $\partial_\mu h_{\text{\tiny(2)}}^{\mu\nu}=0$. Regarding Eqs. \eqref{tau00p1} and \eqref{tau0jp1}, the zeroth component of the conservation equation reveals that
\begin{align}\label{harmonic-p}
\nonumber
0&=\frac{1}{c}\partial_t \prescript{p}{}\tau^{00 \text{\tiny(1)}}_{\text{ eff}}+\partial_j \prescript{p}{}\tau^{0j \text{\tiny(1)}}_{\text{ eff}}\\
& =\partial_t\rho^*+\big(1+\alpha\big)\partial_j\big(\rho^*v^j\big)+O(c^{-2}).
\end{align}
Substituting Eq. \eqref{rhos} in the above relation, we conclude that $\alpha\partial_j\big(\rho^*v^j\big)+O(c^{-2})=0$. This result gives us three possible options which are: I-- $\alpha$ is zero (return to GR). II-- $\partial_j\big(\rho^*v^j\big)$ is divergence-free (a specific type of fluid). III-- The free parameter of the theory is as small as the 1\tiny PN \normalsize correction in GR\footnote{As the terms like $\big(v^2/c^2\big)\partial_j\big(\rho^*v^j\big)$ appear in the GR part of the $O(c^{-2})$ term in Eq. \eqref{harmonic-p}, this case is quite reasonable, cf. Eq. (8.109) of \cite{poisson2014gravity}.} $(\alpha\sim O(c^{-2}))$. Among these cases, the last one is of interest. In this case, we are not in the GR framework, nor is the fluid system necessarily limited to a particular class. In fact, the harmonic gauge condition/conservation statement, Eqs. \eqref{harmonic-gauge} and \eqref{conservation-equation}, together with the baryon number conservation \eqref{rhos} dictate that the magnitude of the free parameter of this theory must be at most of the order of the 1\tiny PN \normalsize terms. 
Henceforth, we treat $\alpha$ as the PN correction and expand the EMPG terms as $O(c^{-2})$. It should be emphasized that we use this rule only for the case $\mathcal{L}_{\text{m}}=p$, and the other case, being a different theory, needs to be examined separately.

Considering this point, we conclude that Eqs. \eqref{h001}-\eqref{hij1} and Eqs. \eqref{tau00p1}-\eqref{tauijp1} reduce to
\begin{subequations}
\begin{align}
\label{h001n}
& \prescript{p}{}h^{00}_{\mathcal{N}\text{\tiny(1)}}=\frac{4}{c^2}U+O(c^{-4}),\\
& \prescript{p}{}h^{0j}_{\mathcal{N}\text{\tiny(1)}}=\frac{4}{c^3}U^j+O(c^{-5}),\\
\label{hij1n}
&\prescript{p}{}h^{ij}_{\mathcal{N}\text{\tiny(1)}}=O(c^{-4}),
\end{align} 
 \end{subequations}
and 
 \begin{subequations}
\begin{align}
\label{tau00p1n}
&\prescript{p}{}\tau^{00 \text{\tiny(1)}}_{\text{ eff}}=\rho^*c^2+O(1),\\
&\prescript{p}{}\tau^{0j \text{\tiny(1)}}_{\text{ eff}}=\rho^*c\,v^j+O(c^{-1}),\\
\label{tauijp1n}
&\prescript{p}{} \tau^{ij \text{\tiny(1)}}_{\text{eff}}=\rho^*v^iv^j+\frac{1}{4\pi G}\Big(\partial^iU\partial^jU-\frac{\delta^{ij}}{2}\partial^nU\partial_nU\Big)\\\nonumber
&~~~~~~~~~~~~~~~~~~~~~~~~~~~~~+\delta^{ij}\Big(p+\alpha\rho^*c^2\Big)+O(c^{-2}),
\end{align} 
 \end{subequations} 
respectively.
According to these terms, the spatial component of $\partial_\mu {\tau_{\text{eff}}^{\mu\nu}}^{(1)}=0$ yields
\begin{align}\label{Euler}
\rho^*\frac{d v^j}{dt}=\rho^*\partial_jU-\partial_jp-\alpha c^2\partial_j\rho^*+O(c^{-2}),
\end{align}
which is the Euler equation in the Newtonian limit of the EMPG theory. Of course, in the Newtonian limit, $\rho^*$ becomes $\rho$.  
In the Sec.~\ref{hij2}, we will use Eqs. \eqref{tau00p1n}-\eqref{tauijp1n} as a source of the wave equation~\eqref{fieldeq_LL} to find the gravitational potential $h^{\mu\nu}_{\text{\tiny (2)}}$.

\subsection{Case $\mathcal{L}_{\text{m}}=-\varepsilon c^2$}

Despite the similarity expected for the two cases $\mathcal{L}_{\text{m}}=p$ and $\mathcal{L}_{\text{m}}=-\varepsilon c^2$ in describing a perfect fluid in GR, the EMPG field relation possesses some differences between these cases that are worth studying. In this section, we then find the source term of the wave equation for the case $\mathcal{L}_{\text{m}}=-\varepsilon c^2$. 

\subsubsection{First iteration}
In a similar method utilized earlier, one can show that
 \begin{subequations}
\begin{align}
\label{T00eff}
&  \prescript{\varepsilon}{}T^{00}_{\text{eff}\text{\tiny(0)}}=(1-\alpha)\rho^*c^2+O(1),\\
&    \prescript{\varepsilon}{}T^{0j}_{\text{eff}\text{\tiny(0)}}=(1-\alpha)\rho^*c\,v^j+O(c^{-1}),\\
\label{Tijeff}
&  \prescript{\varepsilon}{}T^{ij}_{\text{eff}\text{\tiny(0)}}=O(1).
\end{align} 
 \end{subequations}
Even the very first step of the calculations gives us a clue to the deviation from the previous case. Obviously, unlike the case with $\mathcal{L}_{\text{m}}=p$, the EMPG term plays a role in the order $c^{2}$ of the time-time component of $\prescript{\varepsilon}{}T^{\mu\nu}_{\text{eff}}$ while its effects disappear from the space-space component. Using the source terms~\eqref{T00eff}-\eqref{Tijeff}, we obtain the components of $ \prescript{\varepsilon}{}h^{\mu\nu}_{\mathcal{N}\text{\tiny(1)}}$ as
 \begin{subequations}
\begin{align}
\label{h00rho}
&    \prescript{\varepsilon}{}h^{00}_{\mathcal{N}\text{\tiny(1)}}=\frac{4}{c^2}(1-\alpha)U+O(c^{-4}),\\
&    \prescript{\varepsilon}{}h^{0j}_{\mathcal{N}\text{\tiny(1)}}=\frac{4}{c^3}(1-\alpha)U^j+O(c^{-4}),\\
\label{hijrho}
& \prescript{\varepsilon}{}h^{ij}_{\mathcal{N}\text{\tiny(1)}}=O(c^{-4}).
\end{align} 
 \end{subequations}
Keeping in mind that $\prescript{\varepsilon}{}h^{\mu\nu}_{\mathcal{W}\text{\tiny(1)}}=0$ in this stage and substituting the above relations back within Eqs.~\eqref{g00rho}-\eqref{detgrho}, we finally arrive at 
  \begin{subequations}
 \begin{align}
 \label{g00rho1}
 &\prescript{\varepsilon}{}g_{00}^{\text{\tiny(1)}}=-1+\frac{2}{c^2}(1-\alpha)U+O(c^{-4}),\\
 &\prescript{\varepsilon}{}g_{0j}^{\text{\tiny(1)}}=-\frac{4}{c^3}(1-\alpha)U^j+O(c^{-4}),\\
 \label{gijrho1}
 &\prescript{\varepsilon}{}g_{ij}^{\text{\tiny(1)}}=\Big(1+\frac{2}{c^2}(1-\alpha)U\Big)\delta^{ij}+O(c^{-4})\\
 &{(-\prescript{\varepsilon}{}g^{\text{\tiny(1)}})}=1+\frac{4}{c^2}(1-\alpha)U+O(c^{-4}).
 \end{align} 
\end{subequations} 
Now, we have enough information to take another step forward in the iteration method.

\subsubsection{Second iteration}
After using the normalization condition and the definition of the scaled density $\rho^*$, we have 
\begin{subequations}
\begin{align}
&\prescript{\varepsilon}{}\gamma_{\text{\tiny (1)}}=1+\frac{1}{c^2}\big(1-\alpha\big)U+\frac{1}{2}\frac{v^2}{c^2}+O(c^{-4}),\\
&\rho^*=\Big(1+\frac{3}{c^2}\big(1-\alpha\big)U+\frac{1}{2}\frac{v^2}{c^2}+O(c^{-4})\Big)\rho.
\end{align}
\end{subequations}
Here, the standard and EMPG energy-momentum tensors are respectively given by 
\begin{subequations}
\begin{align}
\label{T00rho}
&\prescript{\varepsilon}{}T^{00}_{\text{\tiny (1)}}=\rho^*c^2+O(1),\\
&\prescript{\varepsilon}{}T^{0j}_{\text{\tiny (1)}}=\rho^*c\,v^j+O(c^{-1}),\\
&\prescript{\varepsilon}{}T^{ij}_{\text{\tiny (1)}}=\rho^*v^iv^j+p\,\delta^{ij}+O(c^{-2}),
\end{align}
\end{subequations}
and
\begin{subequations}
\begin{align}
&\prescript{\varepsilon}{}T^{00 \text{\tiny (1)}}_{\text{\tiny EMPG}}=-\alpha\rho^*c^2+O(1),\\
&\prescript{\varepsilon}{}T^{0j\text{\tiny (1)}}_{\text{\tiny  EMPG}}=-\alpha\rho^*c\,v^j+O(c^{-1}),\\
&\prescript{\varepsilon}{}T^{ij\text{\tiny (1)}}_{\text{\tiny  EMPG}}=-\alpha\rho^*v^iv^j+O(c^{-2}).
\end{align}
\end{subequations}
Moreover, one can show that the Landau-Lifshitz and harmonic pseudotensors are simplified as
 \begin{subequations}
\begin{align}
&(-\prescript{\varepsilon}{}g_{\text{\tiny(1)}})\prescript{\varepsilon}{}t^{00 \text{\tiny(1)}}_{\text{ LL}}=O(1),\\
& (-\prescript{\varepsilon}{}g_{\text{\tiny(1)}})\prescript{\varepsilon}{}t^{0j \text{\tiny(1)}}_{\text{ LL}}=O(c^{-1}),\\
&(- \prescript{\varepsilon}{}g_{\text{\tiny(1)}}) \prescript{\varepsilon}{}t^{ij \text{\tiny(1)}}_{\text{ LL}}=\frac{1}{4\pi G}\big(\alpha-1\big)^2\Big(\partial^iU\partial^jU\\\nonumber
&~~~~~~~~~~~~~~~~~~~~~~~~~~~~-\frac{1}{2}\delta^{ij}\partial^nU\partial_nU\Big)+O(c^{-2}),
\end{align} 
 \end{subequations} 
and 
 \begin{subequations}
\begin{align}
\label{t00Hrho}
& (-  \prescript{\varepsilon}{}g_{\text{\tiny(1)}})  \prescript{\varepsilon}{}t^{00 \text{\tiny(1)}}_{\text{ H}}=O(c^{-2}),\\
&(- \prescript{\varepsilon}{}g_{\text{\tiny(1)}}) \prescript{\varepsilon}{}t^{0j \text{\tiny(1)}}_{\text{ H}}=O(c^{-3}),\\
\label{tijHrho}
&  (- \prescript{\varepsilon}{}g_{\text{\tiny(1)}}) \prescript{\varepsilon}{}t^{ij \text{\tiny(1)}}_{\text{H}}=O(c^{-4}),
\end{align} 
 \end{subequations} 
respectively. It should be mentioned that these relations are constructed from Eqs.~\eqref{h00rho}-\eqref{hijrho}. Eqs.~\eqref{t00Hrho}-\eqref{tijHrho} reveal that at this stage of our derivation, the harmonic pseudotensor has no role.
Finally, the components of $ \prescript{\varepsilon}{}\tau^{\mu\nu \text{\tiny(1)}}_{\text{ eff}}$ 
built from Eqs.~\eqref{T00rho}-\eqref{tijHrho} are written as 
 \begin{subequations}
\begin{align}
\label{tau00rho1}
& \prescript{\varepsilon}{}\tau^{00 \text{\tiny(1)}}_{\text{ eff}}=(1-\alpha)\rho^*c^2+O(1),\\
&  \prescript{\varepsilon}{}\tau^{0j \text{\tiny(1)}}_{\text{ eff}}=(1-\alpha)\rho^*c\,v^j+O(c^{-1}),\\
\label{tauijrho1}
&  \prescript{\varepsilon}{}\tau^{ij \text{\tiny(1)}}_{\text{eff}}=(1-\alpha)\rho^*v^iv^j+p\,\delta^{ij}+\frac{1}{4\pi G}\big(1-\alpha\big)^2\\\nonumber
&~~~~~~~~~~~~~~~~~~~~~~~\times\Big(\partial^iU\partial^jU-\frac{1}{2}\delta^{ij}\partial^nU\partial_nU\Big)+O(c^{-2}),
\end{align} 
 \end{subequations}
up to the required degree of accuracy. 

We can now impose the harmonic gauge condition/ conservation equation $\partial_\mu \prescript{\varepsilon}{}\tau^{\mu\nu \text{\tiny(1)}}_{\text{ eff}}=0$. Its zeroth and spatial components respectively show that
\begin{subequations}
\begin{align}
\label{Eq2}
& \big(1-\alpha\big)\Big(\partial_t\rho^*+\partial_j\big(\rho^*v^j\big)\Big)+O(c^{-2})=0\\
\label{Eulerrho}
&\big(1-\alpha\big)\rho^*\frac{dv^j}{dt}=\big(1-\alpha\big)^2\rho^*\partial^jU-\partial^jp+O(c^{-2}).
\end{align}
\end{subequations} 
In the leading order, Eq. \eqref{Eq2} recovers Eq. \eqref{rhos}, and Eq. \eqref{Eulerrho} illustrates the EMPG Euler equation in the Newtonian limit.  
As seen, the harmonic gauge condition, unlike the previous case, does not constrain the magnitude of the free parameter $\alpha$. Therefore, in the case $\mathcal{L}_{\text{m}}=-\varepsilon c^2$, up to this stage of calculation, $\alpha$ are not forced to be as small as the $O(c^{-2})$ terms.

\subsection{$h^{ij}_{\text{\tiny (2)}}$ in the wave zone }\label{hij2}

In the transverse-tracefree gauge imposed in the following, the transverse-tracefree part of the space-space component of the gravitational potential in the wave zone is all we need to study the radiative effects. Therefore, we focus our attention on this component of $h^{\mu\nu}_{\text{\tiny(2)}}$.

\subsubsection{Near-zone portion}

First, we obtain its near-zone portion where the source and field points are located in the near and wave zones, respectively. In the preceding sections, the source terms of this potential are obtained for the two cases $\mathcal{L}_{\text{m}}=p$ and $\mathcal{L}_{\text{m}}=-\varepsilon c^2$. As these cases provide different source terms, it is reasonable to expect that in this theory, GW signals propagate differently and may induce different gravitational effects depending on the choice of the Lagrangian density. In the following, we introduce the general form of $h^{ij}_{\mathcal{N}\text{\tiny (2)}}$ and then specify it for each case.

Using Eq.~\eqref{hNwave}, for the case $l=0$, we have 
\begin{align}
\label{hijN2l=0}
&{h^{ij}_{\mathcal{N}\text{\tiny (2)}}}\mid_{l=0}=\frac{2G}{c^4}\frac{1}{r}\partial_{tt}\int_{\mathfrak{M}}c^{-2}\tau^{00}_{\text{ eff\tiny{(1)}}}x'^ix'^j{\rm d}^3x'\\\nonumber
&+\frac{2G}{c^4}\frac{1}{r}\oint_{\partial\mathfrak{M}}\Big(2\tau^{q(i}_{\text{ eff\tiny{(1)}}}x'^{j)}-\partial_n\tau^{qn}_{\text{ eff\tiny{(1)}}}x'^ix'^j\Big){\rm d}S_q,
\end{align}
where $A^{(i}B^{j)}=1/2\big(A^{i}B^{j}+A^{j}B^{i}\big)$, and the identity 
\begin{align}
&\tau^{ij}_{\text{eff}}=\frac{1}{2}\partial_{00}\big(\tau^{00}_{\text{ eff}}x^ix^j\big)+\frac{1}{2}\partial_q\big(2\tau^{q(i}_{\text{eff}}x^{j)}-\partial_n\tau^{qn}_{\text{eff}}x^ix^j\big)
\end{align}
is applied. This is one of the results of the harmonic gauge condition which can now be imposed  since we are in the last iteration. For more detail, see chapter 7 of~\cite{poisson2014gravity}.  According to the previously mentioned fact, the matter part of the system has no role in the above surface integrals. This is where we lose the effect of the EMPG terms, i.e., $\alpha \rho^* c^2$ in $\prescript{p}{}\tau^{ij \text{\tiny(1)}}_{\text{eff}}$. The only possible contribution to these surface integrals comes from the second and third terms in Eq.~\eqref{tauijp1n} as well as the third and fourth terms in Eq.~\eqref{tauijrho1} in the cases $\mathcal{L}_{\text{m}}=p$ and $\mathcal{L}_{\text{m}}=-\varepsilon c^2$,  respectively. On the other hand, it is shown that the surface integrals constructed from $\partial^iU\partial^jU$ have no role in the potential and they can be omitted freely~\cite{poisson2014gravity}. Thus, we drop the surface integrals in Eq.~\eqref{hijN2l=0}.   

Setting $l=1$ in Eq.~\eqref{hNwave}, we arrive at 
\begin{align}\label{hijN2l=1}
 \nonumber
&{h^{ij}_{\mathcal{N}\text{\tiny (2)}}}\mid_{l=1}=-\frac{2G}{c^4}\\
\nonumber
&\times\partial_{n}\bigg[\frac{1}{r}\partial_t\int_{\mathfrak{M}}\big(2c^{-1}\tau^{0(i}_{\text{ eff\tiny{(1)}}}x'^{j)}x'^n-c^{-1}\tau^{0n}_{\text{ eff\tiny{(1)}}}x'^ix'^j\big){\rm d}^3x'\\
&+\frac{1}{r}\oint_{\partial\mathfrak{M}}\Big(2\tau^{q(i}_{\text{ eff\tiny{(1)}}}x'^{j)}x'^n-\tau^{nq}_{\text{ eff\tiny{(1)}}}x'^ix'^j\Big){\rm d}S_q\bigg].
 \end{align}
It should be mentioned, here, we use the identity
\begin{align}
\nonumber
\tau^{ij}_{\text{eff}}x^p=&\frac{1}{2}\partial_0\Big(2\tau^{0(j}_{\text{ eff}}x^{j)}x^p-\tau^{0p}_{\text{eff}}x^ix^j\Big)\\\label{eq2}
&+\frac{1}{2}\partial_n\Big(2\tau^{n(i}_{\text{eff}}x^{j)}x^p-\tau^{pn}_{\text{eff}}x^ix^j\Big).
\end{align}
In a similar fashion to the previous argument, the surface integrals in Eq.~\eqref{hijN2l=1} can be discarded. Moreover, according to the components of $\prescript{p}{}\tau^{\mu\nu }_{\text{eff}\text{\tiny(1)}}$ and $\prescript{\varepsilon}{}\tau^{\mu\nu}_{\text{eff} \text{\tiny(1)}}$, it can be shown that the volume integrals in this equation are of 0.5\tiny PN \normalsize order smaller than those in Eq.~\eqref{hijN2l=0}. As our goal is to study the leading order of the gravitational potential, we drop this part as well. Finally, we have    
\begin{align}
\label{hijN2}
& h^{ij}_{\mathcal{N}\text{\tiny (2)}}=\frac{2G}{c^4}\frac{1}{r}\partial_{tt}\int_{\mathfrak{M}}c^{-2}\tau^{00}_{\text{ eff\tiny{(1)}}}x'^ix'^j{\rm d}^3x'+O(c^{-5}).
\end{align}

In order to find this potential for the cases $\mathcal{L}_{\text{m}}=p$ and $\mathcal{L}_{\text{m}}=-\varepsilon c^2$, we insert Eqs.~\eqref{tau00p1n} and~\eqref{tau00rho1} into Eq.~\eqref{hijN2}, respectively. So, we get
\begin{align}
\label{hijN2prho}
\prescript{p,\varepsilon}{}h^{ij}_{\mathcal{N}\text{\tiny (2)}}=\frac{2G}{c^4}\frac{\prescript{p,\varepsilon}{}{{\ddot{\mathcal{I}}}^{ij}_{\text{\tiny EMPG}}}}{r}
+O(c^{-5}),
\end{align}
in which
\begin{subequations}
\begin{align}
\label{Iijp}
&\prescript{p}{}{\mathcal{I}^{ij}_{\text{\tiny EMPG}}}=\int_{\mathfrak{M}}\rho^*(\tau,\bm{x}')x'^ix'^j{\rm d}^3x',\\
\label{Iijrho}
&\prescript{\varepsilon}{}{\mathcal{I}^{ij}_{\text{\tiny EMPG}}}=\int_{\mathfrak{M}}\big(1-\alpha\big)\rho^*(\tau,\bm{x}')x'^ix'^j{\rm d}^3x',
\end{align}
\end{subequations}
are the quadrupole-moment tensors written to the leading PN order in this theory for the two different Lagrangian densities.
Here, the overdot shows the derivative with respect to $t$. It is seen that for the model $\eta=1/2$ with $\mathcal{L}_{\text{m}}=p$, the EMPG quadrupole-moment tensor and consequently the EMPG gravitational potential $\prescript{p}{}h^{ij}_{\mathcal{N}}$ are equal to those in GR in this order. On the other hand, for the next case with $\mathcal{L}_{\text{m}}=-\varepsilon c^2$, the difference between this theory and GR manifests itself even in the leading PN order. We will examine this issue in the next section.

\subsubsection{Wave-zone portion}

Now, we turn to find the wave-zone portion of the gravitational potential, $h^{ij}_{\mathcal{W}\text{\tiny(2)}}$. Adding this part to the near-zone one, $h^{ij}_{\mathcal{N}\text{\tiny(2)}}$, found earlier, we actually have enough information to study the radiative effects of gravity. Here, both the source and field points are located in the wave zone. To construct this part of the potential, we need to find its source terms, i.e., the Landau-Lifshitz and harmonic pseudotensors that can exist in the wave zone. To do so, we first introduce their foundations, i.e., $h^{\mu\nu}_{\text{\tiny (1)}}$ in the wave zone. We should be noted that $h^{\mu\nu}_{\text{\tiny (1)}}$ in the near zone has previously been obtained above.

For the case $\mathcal{L}_{\text{m}}=p$, inserting Eqs.~\eqref{T000}-\eqref{Tij0} in the solution~\eqref{hNwave}, we obtain 
\begin{subequations}
\begin{align}
\label{h00N}
&  \prescript{p}{}h^{00}_{\mathcal{N}\text{\tiny(1)}}=\frac{4\,G}{c^2}\frac{1}{r}M_0+O(c^{-4}),\\
&  \prescript{p}{} h^{0j}_{\mathcal{N}\text{\tiny(1)}}=\frac{4\,G}{c^3}\frac{1}{r}P_0^j+O(c^{-5}),\\
\label{hijN}
&\prescript{p}{}h^{ij}_{\mathcal{N}\text{\tiny(1)}}=O(c^{-4}),
\end{align}
\end{subequations}
where
\begin{align}
& M_0=\int_{\mathfrak{M}}\rho^*{\rm d}^3x, ~~~~~~~~~~~~ P_0^j=\int_{\mathfrak{M}}\rho^*v^j {\rm d}^3x.
\end{align}
Here, $M_0$ is interpreted as the total mass inside the near zone. We recall that in this case, $\alpha\sim O(c^{-2})$.  As in this step, there is no source term outside $\mathfrak{M}$, one can conclude that $\prescript{p}{}h^{\mu\nu}_{\mathcal{W}\text{\tiny(1)}}=0$ and consequently  $\prescript{p}{}h^{\mu\nu}_{\text{\tiny(1)}}=\prescript{p}{}h^{\mu\nu}_{\mathcal{N}\text{\tiny(1)}}$. Also, since there is no EMPG correction in Eqs. \eqref{h00N}-\eqref{hijN}, the components of $\prescript{p}{}t^{\mu\nu \text{\tiny(1)}}_{\text{ LL}}$ and $\prescript{p}{}t^{\mu\nu \text{\tiny(1)}}_{\text{ H}}$ would be the same as those obtained in GR.  The rest of the source term~\eqref{tau} is the standard and EMPG energy-momentum tensors. 
On the other hand, for a perfect fluid, these tensors are entirely made up of the matter parts of the system being deep inside the near zone. Keeping these facts in mind, the effective energy-momentum pseudotensor becomes  
\begin{subequations}
\begin{align}
 & \prescript{p}{}\tau^{00 \text{\tiny(1)}}_{\text{eff}}=-\frac{7\,G}{8\,\pi}\frac{1}{r^4}M_0^2+O(c^{-2}),\\
 &\prescript{p}{}\tau^{0j \text{\tiny(1)}}_{\text{eff}}=O(c^{-1}),\\
 \label{eq1}
 &\prescript{p}{}\tau^{ij \text{\tiny(1)}}_{\text{eff}}=\frac{G}{4\pi } \frac{1}{r^4}M_0^2\Big(n^in^j-\frac{1}{2}\delta^{ij}\Big)+O(c^{-2}).
 \end{align} 
  \end{subequations}
Here, $\bm{n}=\bm{x}/r$ is a unit vector representing the direction of the field point $\bm{x}$.  To simplify these terms, Eq.~\eqref{rhos} is also utilized.
It is worth noting that both the GR and EMPG terms contribute to the order $c^{-2}$ in the time-time and space-space components of $\prescript{p}{}\tau^{\mu\nu \text{\tiny(1)}}_{\text{eff}}$. As the source of $\prescript{p}{}h^{ij}_{\mathcal{W}\text{\tiny(2)}}$, Eq. \eqref{eq1}, is similar to the GR case, we recover the GR result
\begin{align}
\label{hij_w2_p}
\prescript{p}{} h^{ij}_{\mathcal{W}\text{\tiny (2)}}=\frac{G^2}{c^4}\frac{1}{r^2}M_0^2\Big(n^{\langle ij \rangle }+\frac{1}{3}\delta^{ij}\Big).
\end{align}
Here, $n^{i}n^{j}=n^{\langle ij \rangle}+\frac{1}{3}\delta^{ij}$ where $n^{\langle ij \rangle }$ is an angular symmetric tracefree (STF) tensor introduced in Eq.~(1.154) of~\cite{poisson2014gravity}.

Before we examine the role of this portion in the total gravitational potential $\prescript{p}{} h^{ij}_{\text{\tiny (2)}}$, let us also find the wave-zone portion for the case $\mathcal{L}_{\text{m}}=-\varepsilon c^2$. In a similar fashion to the previous part, we first obtain the components of $\prescript{\varepsilon}{}h^{\mu\nu}_{\mathcal{N}\text{\tiny(1)}}$ as
\begin{subequations}
\begin{align}
&    \prescript{\varepsilon}{}h^{00}_{\mathcal{N}\text{\tiny(1)}}=\frac{4\,G}{c^2}\big(1-\alpha\big)\frac{1}{r}M_0+O(c^{-3}),\\
&    \prescript{\varepsilon}{}h^{0j}_{\mathcal{N}\text{\tiny(1)}}=\frac{4\,G}{c^3}(1-\alpha)\frac{1}{r}P_0^j+O(c^{-4}),\\
&  \prescript{\varepsilon}{}h^{ij}_{\mathcal{N}\text{\tiny(1)}}=O(c^{-4}),
\end{align}
\end{subequations}
where Eqs.~\eqref{T00eff}-\eqref{Tijeff} are used for the source terms. Using these, we then find that
 \begin{subequations}
\begin{align}
&(- \prescript{\varepsilon}{}g_{\text{\tiny(1)}}) \prescript{\varepsilon}{}t^{00 \text{\tiny(1)}}_{\text{ LL}}=-\frac{7\,G}{8\,\pi}\big(1-\alpha\big)^2\frac{1}{r^4}M_0^2+O(c^{-2}),\\
&(-\prescript{\varepsilon}{}g_{\text{\tiny(1)}})\prescript{\varepsilon}{}t^{0j \text{\tiny(1)}}_{\text{ LL}}=O(c^{-1}),\\\nonumber
&(-\prescript{\varepsilon}{}g_{\text{\tiny(1)}})\prescript{\varepsilon}{}t^{ij \text{\tiny(1)}}_{\text{ LL}}=\frac{G}{4\pi }\big(1-\alpha\big)^2\frac{1}{r^4}M_0^2\Big(n^in^j-\frac{1}{2}\delta^{ij}\Big)\\\label{eq3}
&~~~~~~~~~~~~~~~~~~~~~+O(c^{-2}),
\end{align} 
 \end{subequations}
as well as
 \begin{subequations}
\begin{align}
& (-\prescript{\varepsilon}{}g_{\text{\tiny(1)}})\prescript{\varepsilon}{}t^{00 \text{\tiny(1)}}_{\text{ H}}=O(c^{-2}),\\
& (- \prescript{\varepsilon}{}g_{\text{\tiny(1)}})  \prescript{\varepsilon}{}t^{0j \text{\tiny(1)}}_{\text{ H}}=O(c^{-3}),\\
\label{eq4}
&  (- \prescript{\varepsilon}{}g_{\text{\tiny(1)}})  \prescript{\varepsilon}{}t^{ij \text{\tiny(1)}}_{\text{H}}=O(c^{-4}).
\end{align} 
 \end{subequations}
As seen, the harmonic pseudotensor has no contribution in the PN order required for the effective energy-momentum pseudotensor. Next, we conclude that 
\begin{align}\label{source-w}
&\prescript{\varepsilon}{}\tau^{ij \text{\tiny(1)}}_{\text{eff}}=\frac{G}{4\pi }\big(1-\alpha\big)^2\frac{1}{r^4}M_0^2\Big(n^{\langle ij \rangle}-\frac{1}{6}\delta^{ij}\Big)+O(c^{-2}).
\end{align}
Here, toward finding $h^{ij}_{\mathcal{W}\text{\tiny(2)}}$, we rewrite its source as shown in Eq.~\eqref{eq7}.
Inserting Eq.~\eqref{source-w} into Eq.~\eqref{eq8} finally reveals that
\begin{align}
\label{hij_W2_rho}
\prescript{\varepsilon}{}h^{ij}_{\mathcal{W}\text{\tiny (2)}}=\frac{G^2}{c^4}\big(1-\alpha\big)^2\frac{1}{r^2}M_0^2\Big(n^{\langle ij \rangle}+\frac{1}{3}\delta^{ij}\Big).
\end{align}
It should be mentioned, to achieve this result, it is assumed that the free parameter of the EMPG theory is constant.

It is seen that depending on the Lagrangian density, $h^{ij}_{\mathcal{W}\text{\tiny (2)}}$ is different; and only after dropping the EMPG parts, Eqs.~\eqref{hij_w2_p} and~\eqref{hij_W2_rho} will be equal. Furthermore, these relations indicate that  $h^{ij}_{\mathcal{W}\text{\tiny (2)}}$ falls off as $r^{-2}$. Nevertheless, the near-zone portion is a linear function of $r^{-1}$, cf. Eq.~\eqref{hijN2prho}. So, for both cases, the wave-zone portion falls off faster than the near-zone one. 
Since our goal is to investigate the radiative effects that are only significant far away form the gravitational system, we set aside  $h^{ij}_{\mathcal{W}\text{\tiny (2)}}$ in comparison with $h^{ij}_{\mathcal{N}\text{\tiny (2)}}$. We then have $h^{ij}_{\text{\tiny (2)}}=h^{ij}_{\mathcal{N}\text{\tiny (2)}}$. This is also the case in GR. In fact, according to the role of the EMPG terms in Eqs.~\eqref{eq1},~\eqref{eq3}, and~\eqref{eq4} and its similarity to the GR one, this fact was a predictable outcome. However, for the sake of completeness, this part of calculations is added in detail here.

To sum up, utilizing Eq.~\eqref{hijN2prho}, one can study the gravitational potential in the wave zone.
For the case $\mathcal{L}_{\text{m}}=p$, given the quadrupole-moment tensor~\eqref{Iijp}, this potential is indeed equal to GR up to the PN order considered here.
It should be mentioned that in~\cite{2022PhRvD.105d4014N}, it is shown that even in the leading PN order, the gravitational potential in the quadratic-EMSG model (EMPG with $\eta=1$) is different from that in GR. Therefore, as expected, the mathematical form of the gravitational potential in the wave zone strongly depends on the value of $\eta$ and as a result the GW signals would behave differently in each EMPG model.   
Here, the GW signals seem to behave similarly in GR and EMPG with $\eta=1/2$ and $\mathcal{L}_{\text{m}}=p$.  On the other hand, in the case $\mathcal{L}_{\text{m}}=-\varepsilon c^2$, the EMPG terms play a role in the gravitational potential even in the leading order. See Eq.~\eqref{Iijrho}. It means that gravitational systems radiate different GWs in the EMPG theory compared to those in GR.    
In the following section, we focus our attention on this fact and attempt to examine GW signals from compact binary systems and their radiative effects in the scale-independent EMSG.

\section{Gravitational-wave radiation}\label{sec3}

We consider a compact binary system as a source of GW signals. For future calculations, let us first introduce the coordinate system. We choose the orbit-adapted frame $(x, y, z)$ whose origin is located at the system's barycenter. The $x$-$y$ plane coincides with the orbital plane so that the $x$-, $y$-, and $z$-axes are aligned with the orbit's major axis, minor axis, and the angular-momentum vector, respectively. In this coordinate system, the bases of the orbital plane are given by $\bm{N}=\big[\cos\varphi, \sin\varphi, 0\big]$ and $\bm{\lambda}=\big[-\sin\varphi, \cos\varphi, 0\big]$ where $\varphi$ is the angle from the orbit's major axis, i.e., $x$-axis.

Also, to simplify the gravitational potential $h^{ij}$ even further, we implement the transverse-tracefree gauge, called TT gauge, which is achieved in the far-away wave zone. It should be noted that as EMSG and its subclasses are not different from GR in the vacuum, the number of polarization modes of GWs does not change. So, in this model, like GR, we will have two usual plus and cross polarizations.

\subsection{Gravitational-wave field and polarizations}

It is assumed that in the binary system, the center of mass of the first and second bodies with masses $m_1$  and $m_2$ are located at  $R_1(t)$ and $R_2(t)$ relative to the system's barycenter, respectively. 
One can straightforwardly show that in the chosen coordinate system, after imposing TT gauge, the STF pieces of the quadrupole-moment tensors~\eqref{Iijp} and~\eqref{Iijrho}, i.e., $\mathcal{I}^{\langle ij\rangle}=\mathcal{I}^{ij}-\frac{1}{3}\delta^{ij}\mathcal{I}^{qq}$,  are given by
\begin{subequations}
\begin{align}
\label{Ip}
&\prescript{p}{}{\mathcal{I}}^{\langle ij\rangle}=\nu\,m\Big(R^iR^j-\frac{1}{3}\delta^{ij}R^2\Big),\\
\label{Irho}
&\prescript{\varepsilon}{}{\mathcal{I}}^{\langle ij\rangle}=\big(1-\alpha\big)\nu\,m\Big(R^iR^j-\frac{1}{3}\delta^{ij}R^2\Big),
\end{align}
\end{subequations}
where $m=m_1+m_2$ is the total mass, $\nu=m_1m_2/m^2$ is the symmetric mass ratio of the system, and $\bm{R}=\bm{R}_1-\bm{R}_2$ measures the separation between two bodies. Here, $R=|\bm{R}|$. As the intrinsic moments of the compact bodies do not play a role in the following calculations, we drop them from the relations \eqref{Ip} and \eqref{Irho}. 
To proceed further, we should find the second time derivative of these quadrupole-moment tensors. For instance, for the case $\mathcal{L}_{\text{m}}=-\varepsilon c^2$, we have $\prescript{\varepsilon}{}{\ddot{\mathcal{I}}}^{ij}=\big(1-\alpha\big)\nu\,m\big( 2v^iv^j+R^ia^j+R^ja^i\big)$ in which $\bm{v}=\dot{\bm{R}}$ and $\bm{a}=\dot{\bm{v}}$ are the relative velocity and acceleration vectors, respectively. Therefore, in this step, the Newtonian description of the orbital
motion, i.e., $\bm{v}$, $\bm{a}$, and $\bm{R}$, in EMPG should be obtained. In fact, as our goal is to study GWs to the leading order, it is sufficient to derive the orbital motion in the Newtonian limit of the scale-independent EMSG.

To do so, we utilize the Euler equation in the Newtonian limit of the theory introduced before.  We define the inertial mass and the center of mass of each body as
\begin{subequations}
\begin{align}
&m_{1}=\int_{V_1} \rho^*\, {\rm d}^3x,\\
&\bm{R}_{1}=m_1^{-1}\int_{V_1}\rho^*\,\bm{x}\,{\rm d}^3x,
\end{align}
\end{subequations}
respectively. Using Eq. \eqref{rhos}, one can show that
\begin{subequations}
\begin{align}
&\frac{dm_1}{dt}=0,\\
& \bm{v}_{1}=\dot{\bm{R}}_1=m_1^{-1}\int_{V_1}\rho^*\,\bm{v} {\rm d}^3x,\\
\label{aA}
&\bm{a}_{1}=\dot{\bm{v}}_1=m_A^{-1}\int_{V_1}\rho^*\,\frac{d\bm{v}}{dt}\,{\rm d}^3x,
\end{align}
\end{subequations}
after considering that there is no flux of matter from the body.
Inserting Eqs. \eqref{Euler} and \eqref{Eulerrho} into Eq. \eqref{aA}, assuming that the pressure/density is zero on the surface of each body, taking bodies being nearly spherical, and neglecting the terms proportional to multipole moments of bodies, we obtain the equation of motion for the body ``1'' as $\prescript{p}{}{\bm{a}}_1=-\big(G\,m_2/R^2\big)\bm{N}$ and $\prescript{\varepsilon}{}{\bm{a}}_1=-\big(1-\alpha)\big(G\,m_2/R^2\big)\bm{N}$, respectively. Here, $\bm{N}=\bm{R}/R$. Given these results, we finally arrive at 
\begin{subequations}
\begin{align}
\label{ap}
&\prescript{p}{}{\bm{a}}=-\frac{G\,m}{R^2}\bm{N},\\
\label{arho}
&\prescript{\varepsilon}{}{\bm{a}}=-\big(1-\alpha\big)\frac{G\,m}{R^2}\bm{N},
\end{align} 
\end{subequations} 
for the EMPG relative acceleration vectors in the binary system.

As shown in Eq. \eqref{ap}, for the case $\mathcal{L}_{\text{m}}=p$, the EMPG correction plays no role in the motion of the body at least in the Newtonian limit. 
So, it turns out that the EMPG effects with $\mathcal{L}_{\text{m}}=p$ do not appear in the leading order of the quadrupole formula for the GW field \eqref{hijN2prho}. To see the possible effects of the scale-independent EMSG with this choice of the matter Lagrangian density, higher PN corrections to the mass quadrupole-moment tensor \eqref{Iijp}, as well as to the orbital motion (in the case of a binary system) should be taken into account. At the same time, one should go beyond the quadrupole formula and obtain higher PN corrections, coming from the radiative multipole moments, to the gravitational potential.
Therefore, in this case, up to the leading order studied here, the EMPG gravitational potential $\prescript{p}{}h^{ij}_{\mathcal{N}}$ does not deviate from that given in GR and further corrections should be investigated in this regard.

On the other hand, for the case $\mathcal{L}_{\text{m}}=-\varepsilon c^2$, not only the definition \eqref{Iijrho} modifies, but also the equation of motion \eqref{arho} is affected by the EMPG corrections. Since in the present work, we focus our attention on the leading PN order and aim to limit the free parameter of this theory to this order, we continue our study with the case $\mathcal{L}_{\text{m}}=-\varepsilon c^2$ whose footprints appear in the leading order of the quadrupole formula, leaving the case $\mathcal{L}_{\text{m}}=p$, which requires tedious computations in higher PN orders, for the future.

It is obvious that by replacing $\big(1-\alpha\big)m$ with $m$, Eq. \eqref{arho} reduces to those in GR. In fact, only the mass is rescaled as $\big(1-\alpha\big)m$ in the equations of motion of the binary system in the Newtonian limit of EMPG. This fact is consistent with the result of~\cite{2022arXiv221004668A}. So, applying this, one can utilize the Keplerian descriptions       
\begin{subequations}
\begin{align}
\label{Rp}
&R=\big(1-\alpha\big)^{-1}\frac{l}{1+e \cos \varphi},\\
\label{dphip}
&\dot{\varphi}=\big(1-\alpha\big)^2\Big(\frac{G\,m}{l^3}\Big)^{\frac{1}{2}}\big(1+e\cos\varphi\big)^2,
\end{align}
\end{subequations}
where $l=h^2/(G\, m)$ is the semi-latus rectum in which $h$ is a constant of the motion, and $e$ is the eccentricity of the orbit. Moreover, in this framework, the total energy contained in the binary system $E$ and its orbital period $P$ are defined as $E=-(1-\alpha)\,G\, m\,\mu/2a$ and $P=2\pi a^{3/2}(G\,m)^{-1/2}(1-\alpha)^{-1/2}$, respectively.  Here, $\mu=m_1m_2/m$ is the reduced mass and  $a=l(1-e^2)^{-1}(1-\alpha)^{-1}$ is the semi-major axis.

Using Eqs.~\eqref{arho},~\eqref{Rp}, and~\eqref{dphip} in the definition of $\prescript{\varepsilon}{}{\ddot{\mathcal{I}}}^{ij}$ \footnote{In~\cite{2022PhRvD.105d4014N}, the Newtonian dynamics of the quadratic-EMSG theory is not considered in the derivation of ${\ddot{\mathcal{I}}}^{ij}$ and the GR version is simply utilized. However, as mentioned in this work, the Newtonian limit of the theory can affect the results and should be taken into account in the calculations.}, after some simplifications, we obtain 
\begin{align}
\nonumber
&\prescript{\varepsilon}{}{\ddot{\mathcal{I}}}^{ij}=\frac{2\,G\,\nu\, m^2}{l}\big(1-\alpha\big)^3\Big[-\big(1+e\,\cos\varphi\\
\nonumber
&-e^2\sin^2\varphi\big)N^iN^j+e\,\sin\varphi\big(1+e\,\cos\varphi\big)\big(N^i\lambda^j+N^j\lambda^i\big)
\\
&+\big(1+e\,\cos\varphi\big)^2\lambda^i\lambda^j\Big].
\end{align}
Finally, regarding the above relation, we reach 
\begin{align}\label{hijp2}
\nonumber
&\prescript{\varepsilon}{}h^{ij}=\frac{4\,G^2}{c^4}\frac{\nu\,m^2}{l}\frac{1}{r}\big(1-\alpha\big)^3\Big[-\big(1+e\,\cos\varphi\\\nonumber
&-e^2\sin^2\varphi\big)N^iN^j+e\,\sin\varphi\big(1+e\,\cos\varphi\big)\big(N^i\lambda^j+N^j\lambda^i\big)\\
&+\big(1+e\,\cos\varphi\big)^2\lambda^i\lambda^j\Big]
\end{align}
for the gravitational potential in the wave zone.\footnote{It is worth mentioning that for the specific value of the free parameter, i.e., $\alpha=1$, the second time derivative of the quadrupole-moment tensor and consequently the GW field would vanish, cf. Eqs.~\eqref{hijp2}. So, one can conclude that for this value of $\alpha$, the radiative aspects of the scale-independent EMSG binary systems would not manifest themselves up to the leading PN order. In other words, for this particular model, i.e., the EMPG theory with $\eta=1/2$ and $\alpha=1$, the EMPG correction prevents binary systems from radiating GWs to this PN order. One then needs to take into account higher PN orders to study GWs in this model.
 On the other hand, in the following, by studying GW observations, we illustrate that the free parameter of the theory  should be placed in the very small interval $\mid \alpha \mid<10^{-5}$ and the value $\alpha=1$ is indeed ruled out. Nonetheless, the mathematical fact saying that GWs do no propagate up to the leading PN order of this EMPG model can be of interest. As can be seen, the EMPG manifests itself in the cubic term} $\big(1-\alpha\big)^3$, which does nothing but rescale the GR estimations; and by dropping $\alpha$, this coefficient reduces to unity and recovers GR. It is worth mentioning that following the above calculation, in the case $\mathcal{L}_{\text{m}}=-\varepsilon c^2$, a linear portion comes from the relativistic aspects of the scale-independent EMSG while the Newtonian dynamics of this theory brings an extra quadratic portion to the GW field.

The final task toward finding the GW signals is to obtain the polarizations $h_{+}$ and $h_{\times}$. To find these components, we should introduce the detector-adapted frame. We exhibit this frame with $(X, Y, Z)$ and assume that the $Z$-axis is oriented towards the direction of the detector and the $X$-$Y$ plane is the sky plane. In fact, the $X$-$Y$ plane is the transverse subspace that is orthogonal to the direction of the GW propagation.
The origin of this frame coincides with the system's barycenter, and the $X$-axis is the intersection between the sky and orbital planes. In this case, we assume that the longitude of the ascending node is zero.
In terms of the bases of this transverse subspace, i.e., $\bm{e}_X$ and $\bm{e}_Y$, the polarizations are obtained as follows
\begin{subequations}
\begin{eqnarray}
\label{h_plus1}
&& h_+=\frac{1}{2}\big(e^j_X e^k_X-e^j_Y e^k_Y\big)h_{jk},\\
\label{h_cross1}
&& h_{\times}=\frac{1}{2}\big(e^j_X e^k_Y+e^j_Y e^k_X\big)h_{jk}.
\end{eqnarray}
\end{subequations}
In this case, one can rewrite $\bm{N}$ and $\bm{\lambda}$ as
\begin{subequations}
\begin{eqnarray}
\nonumber
&&\bm{N}=\left[\cos\left(\omega+\varphi\right),\cos\iota\sin\left(\omega+\varphi\right),\sin\iota\sin\left(\omega+\varphi\right)\right],\\\label{n}
\\
\nonumber
&&\bm{\lambda}=\left[-\sin\left(\omega+\varphi\right),\cos\iota\cos\left(\omega+\varphi\right),\sin\iota\cos\left(\omega+\varphi\right)\right],\\\label{L}
\end{eqnarray}
\end{subequations}
in terms of the detector-adapted bases, respectively.
Here, $\iota$ displays the angle between the orbital and sky planes and $\omega$ represents the angle between the $x$ and $X$ axes.
Now, by applying Eqs.~\eqref{hijp2} and~\eqref{h_plus1}-\eqref{h_cross1}, and also using these bases, i.e., equations~\eqref{n}-\eqref{L}, we derive that 
\begin{eqnarray}
\label{h}
\prescript{\varepsilon}{}h_{+}=\prescript{\varepsilon}{}h_{0}H_{+},~~~~~~~~~~~\prescript{\varepsilon}{}h_{\times}=\prescript{\varepsilon}{}h_{0}H_{\times},
\end{eqnarray}
where the GW amplitude is given by
\begin{align}
\label{h0p}
&\prescript{\varepsilon}{} h_{0}=\frac{2 G^2}{c^4}\frac{\nu\, m^2}{ l\, r}\big(1-3\alpha\big).
\end{align}
Moreover, 
\begin{eqnarray}
\nonumber
&& H_{+}= -\left(1+\cos^2\iota\right)\bigg[\cos\left(2\varphi+2\omega\right)+\frac{5}{4}e\cos\left(\varphi+2\omega\right)\\\nonumber
&&+\frac{1}{4}e\cos\left(3\varphi+2\omega\right)+\frac{1}{2}e^2\cos 2\omega\bigg]+\frac{1}{2}e\sin^2\iota \left(\cos\varphi+e\right),\\
\end{eqnarray}
and
\begin{eqnarray}
\nonumber
&& H_{\times}=-2\cos\iota\bigg[\sin\left(2\varphi+2\omega\right)+\frac{5}{4}e\sin\left(\varphi+2\omega\right)\\
&&+\frac{1}{4}e\sin\left(3\varphi+2\omega\right)+\frac{1}{2}e^2\sin 2\omega\bigg],
\end{eqnarray}
represent the plus and cross scale-free polarizations, respectively. Therefore, up to this order, the scale-independent EMSG with $\mathcal{L}_{\text{m}}=-\varepsilon c^2$ only modifies the wave amplitude, leaving the scale-free polarizations unchanged. 
It should be pointed out that to obtain the above results, we consider that the free parameter of the scale-independent EMSG is small. It is actually a reasonable assumption, because the EMPG would otherwise change the solar-system weak-field tests of gravity dramatically. 

\subsection{Radiative losses}\label{Radiative losses}

In this part, we investigate the rate at which GW radiations remove energy from their gravitating sources, e.g., compact binary systems. In fact, according to the change of $h^{ij}$ in this theory, the system may lose energy to modified gravitational radiation in EMPG. To show this, we utilize the following famous quadrupole relation
\begin{align}
\mathcal{P}=\frac{G}{5c^5}\dddot{\mathcal{I}}^{\langle ij\rangle}\dddot{\mathcal{I}}_{\langle ij\rangle}, 
\end{align}
where $\mathcal{P}$ is the flux of gravitational energy in the far-away wave zone. Here, the TT gauge is imposed. The rest of this section is devoted to extracting this flux in the scale-independent EMSG theory.

In order to find the gravitational energy flux, we should have enough information about the third time derivative of the STF piece of the quadrupole-moment tensor of the matter distribution in the scale-independent EMSG, i.e.,  we should obtain
\begin{align}\label{Iddd}
\nonumber
&\prescript{\varepsilon}{} {\dddot{\mathcal{I}}}^{\langle ij\rangle}=\nu\,m\,\big(1-\alpha\big)\Big[3\big(v^ia^j+v^ja^i\big)+R^i\dot{a}^j+R^j\dot{a}^i\\
&~~~~~~~~~~~~~~~~~~~~~~~~~~-\frac{2}{3}\delta^{ij}\big(3v a+R\dot{a}\big)\Big].
\end{align}
Regarding the scale-independent EMSG Newtonian dynamics previously obtained, one can easily reach
\begin{equation}
\begin{aligned}\label{adp}
\dot{\bm{a}}=&\big(1-\alpha\big)^5\Big(\frac{G^3\,m^3}{l^7}\Big)^{\frac{1}{2}}\big(1+e\,\cos\varphi\big)^3\\
&~~~~~~~~~~~~~~~\times\Big[2\,e\,\sin\varphi\,\bm{N}-\big(1+e\,\cos\varphi\big)\bm{\lambda}\Big].
\end{aligned}    
\end{equation}
Making the substitution Eqs.~\eqref{arho},~\eqref{Rp},~\eqref{dphip}, and~\eqref{adp} into Eq.~\eqref{Iddd}, after some manipulations, and inserting the result into the definition of the energy flux, we finally reach
\begin{equation}
\begin{aligned}\label{flux_p}
\prescript{\varepsilon}{}{\mathcal{P}}=&\frac{32}{5}\frac{\nu^2}{G}\big(1-\alpha\big)^{10}\Big(\frac{G\, m}{c\,l}\Big)^5\big(1+e\,\cos\varphi\big)^4\\
&\times\Big[1+2\,e\,\cos\varphi+\frac{1}{12}e^2\big(1+11\cos^2\varphi\big)\Big].
\end{aligned}
\end{equation}

This result leads to the modification of the relevant post-Keplerian parameter, i.e., the first time derivative of the orbital period of binary systems, in this modified theory of gravity.  
To obtain this parameter, we apply the energy-balance equation in the averaged form 
\begin{align}
\dot{E}=-\langle\mathcal{P}\rangle, 
\end{align}
where $\langle\mathcal{P}\rangle$ is the orbital average of the flux of gravitational energy which is defined as $\langle\mathcal{P}\rangle=(2\pi)^{-1}\big(1-e^2\big)^{3/2}\int_{0}^{2\pi}\big(1+e \cos\varphi\big)^{-2}\mathcal{P}(\varphi){\rm d}\varphi$.
Regarding the relation between the energy and the orbital period and using Eqs.~\eqref{flux_p}, we finally arrive at 
\begin{equation}
\begin{aligned}\label{Tdotp}
{\dot{P}}_{\text{\tiny EMPG}}=&-\frac{192\pi}{5}\Big(\frac{G \mathcal{M}}{c^3}\frac{2\pi}{P}\Big)^{\frac{5}{3}}\Big(1-\frac{8}{3}\alpha\Big)\\
&\times\big(1-e^2)^{-\frac{7}{2}}\Big[1+\frac{73}{24}e^2+\frac{37}{96}e^4\Big],
\end{aligned}    
\end{equation}
for the first time derivative of the orbital period in EMPG. Here, $\mathcal{M}=\nu^{3/5}m$ is the chirp mass.
This relation exhibits the extra EMPG portion in GWs radiation, and consequently its effects in extracting orbital energy from the binary motion, which leads to the secular change of the corresponding Keplerian parameter, i.e., the orbital period. It should be noted that in order to obtain the secular change of other Keplerian parameters, one should also study the momentum-balance equation. However, as our aim is to test this theory using the best-measured post-Keplerian parameter, $\dot{P}$, in the current work, we restrict ourselves to these results and leave further study for the future.

\section{Tests from gravitational-wave and binary-pulsar tests}\label{sec4}

Applying the results obtained in the previous section together with GW observations from the relativistic compact binary systems, we attempt to constrain the scale-independent EMSG theory with $\mathcal{L}_{\text{m}}=-\varepsilon c^2$. Here, two different GW observations are applied, the direct and indirect observations. The direct ones are those that are provided by the GW observatories like LIGO and Virgo. This type of observation indeed determines the actual GW signals far in the wave zone of sources. On the other hand, the indirect observations are referred to the binary pulsar experiments which reveal the influence of the GW propagation on the intrinsic parameters of its generator such as the orbital period of the binary system. So, this type of observation indicates GW effects in the near zone of binary systems.

\subsection{Tests from GW observations of binaries}

The direct observation of the GW signals of binary systems are highly sensitive to the change of the frequency and the phase of GWs. When a binary system is in the inspiral phase, the GW signals can inter the detector's sensitive bandwidth. Therefore, practically, hundreds to tens of thousands of cycles before the merger can be observed. The total accumulated GW phase over the cycles in the bandwidth of detectors is given by
\begin{align}\label{GW_phase}
\Phi_{\text{GW}}=\int_{f_{\text{in}}}^{f_{\text{out}}}2\pi\,(f/\dot{f}){\rm d}f,
\end{align}
where $f$ is the GW frequency and $f_{\text{in}}$ and $f_{\text{out}}$ stand for the frequency at which the GW signal enters and leaves the detector's bandwidth, respectively. The theoretical GW template would be accurate compared to the actual signal if a change in $\Phi_{\text{GW}}$ be smaller than $\pi$ radians~\cite{1994PhRvD..50.6058W}. We use this fact to constrain the free parameter of the scale-independent EMSG. It should be mentioned that \cite{1994PhRvD..50.6058W} introduces this method  to find a bound on the coupling constant $\omega_{\text{BD}}$ of Brans-Dicke theory in the strong-field regime.      

Here, we consider that the orbit is circular. Regarding the GW polarizations obtained before, one can deduce that $f=\Omega/\pi$ for this case where $\Omega$ is the orbital frequency given by $\Omega=2\pi/P$.  
Bearing this fact in mind and also utilizing Eq.~\eqref{Tdotp}, it turns out that the GW frequency evolves as 
\begin{align}\label{fdotp}
{\dot{f}}=\frac{96\pi}{5}\Big(\frac{\pi G  \mathcal{M}}{c^3}\Big)^{\frac{5}{3}}\Big(1-\frac{8}{3}\alpha\Big)f^{\frac{11}{3}}.
\end{align}
Insertion of this relation into Eq.~\eqref{GW_phase} reveals that
\begin{align}
\Phi_{\text{GW}}=-\frac{1}{16}\Big(\frac{\pi G  \mathcal{M}}{c^3}\Big)^{-\frac{5}{3}}\Big(1+\frac{8}{3}\alpha\Big)\Big(f^{-\frac{5}{3}}_{\text{out}}-f^{-\frac{5}{3}}_{\text{in}}\Big).
\end{align}
It is seen that the scale-independent EMSG can change the total accumulated GW phase. Using the phase-shift estimation mentioned above and demanding that this change induced by the scale-independent EMSG terms should be less than $\pi$, we find the following limit 
\begin{align}
\arrowvert{\alpha}\arrowvert<5.2\times10^{-5}\Big(\frac{\mathcal{M}}{M_{\odot}}\Big)^{\frac{5}{3}}\Big(\frac{f_{\text{in}}}{30\,\text{Hz}}\Big)^{\frac{5}{3}},
\end{align}
for the magnitude of the free parameter of the theory in the case $\mathcal{L}_{\text{m}}=-\varepsilon c^2$. It should be mentioned that as $f_{\text{out}}$ is of the order $1000\,\text{Hz}$, two orders of magnitude larger than $f_{\text{in}}$, we drop the term $f^{-\frac{5}{3}}_{\text{out}}$ in the above derivation.\footnote{ For instance, for the ground-based LIGO and Virgo observatories, $f_{\text{in}}$ is of the order $10\,\text{Hz}$.} 

Given the above result, the smaller the chirp mass, the tighter the limits are.
In other words, among the neutron star--neutron star, neutron star--black hole, black hole--black hole binary systems, the first system would set the strongest limit on the magnitude of the scale-independent EMSG free parameter. Regarding this point, among the 181 recent events listed at \url{https://gw-openscience.org}, we select those with the smallest chirp mass, e.g., GW190425 with $\mathcal{M}=(1.44\pm 0.02)\,M_{\odot}$ and GW170817 with $\mathcal{M}=(1.186\pm 0.001)\,M_{\odot}$.
From these events, the tightest bounds can be obtained as follows:  $\arrowvert {\alpha}\arrowvert<4.7\times10^{-5}$ and $\arrowvert{\alpha}\arrowvert<4.5\times10^{-5}$ for  GW190425 and GW170817, respectively. 
Here, we take $f_{\text{in}}=19.4\,\text{Hz}$ for GW190425 and $f_{\text{in}}=23\,\text{Hz}$  for GW170817 following~\cite{2020ApJ...892L...3A,2020CQGra..37d5006A}. In fact, by decreasing the chirp mass, more cycles of binary systems could happen in the frequency band of the detectors, and as a result, more the scale-independent EMSG effects would accumulate in the total GW phase. For instance, for GW190425 and GW170817, there were $\sim 3900$ and $\sim3000$ cycles, respectively, before the merger~\cite{2020ApJ...892L...3A,2017PhRvL.119p1101A}. Therefore, it is reasonable to expect that a stronger limit would be achieved from these events.

Another possible way to constrain the free parameter of the theory is to analyze the GW waveform. As it is shown earlier, the new terms contribute to the GW amplitude, leaving the scale-free plus and cross polarizations unchanged. Let us rewrite the modified amplitude as ${h}_0={h}_0+\Delta$ where $\Delta\propto{h}_0\times{\alpha}$ is the scale-independent EMSG correction. This new correction should in principle be greater than the error in estimating ${h}_0$ in order to be detected or even limited. \cite{1994PhRvD..50.6058W} shows that the root mean square error for $\ln {h}_0$ is equal to the inverse of the signal-to-noise ratio of a given signal, i.e., $\Delta(\ln {h}_0)=1/(S/N)$. According to the information provided at \url{https://gw-openscience.org}, the event GW170817 with $S/N=33$ has the highest signal-to-noise ratio among other recent events. This means that in the best case, the minimum value of $\alpha$ which can be detected by applying the GW amplitude analysis is of the order $10^{-2}$.  Therefore, although the scale-independent EMSG plays a role in the GW amplitude, compared to the phase-shift method which is sensitive to $\alpha$ of the order $10^{-5}$, the GW amplitude analysis does not impose a tight restriction and the scale-independent EMSG modifications may get lost in it.

It should be noticed that to obtain the above constraints, it is assumed that the chirp mass $\mathcal{M}$ is already known. In fact, during this calculation, we consider that GR is the valid gravity theory, as a priori expectation, and then attempt to set a bound on $\alpha$. However, let us define the chirp mass in the scale-independent EMSG as $\mathcal{M}_{\text{EMPG}}=\mathcal{M} \big(1-8\alpha/5\big)$. Regarding this rescaled chirp mass, one can see that Eqs.~\eqref{Tdotp} and~\eqref{fdotp} reduce identically to those in GR up to the leading PN order. On the other hand, since the chirp mass of the binary systems are measured purely via GW observations, one cannot indeed measure this rescaling and consequently the value of $\alpha$ is utterly undetectable. Thus, from a practical perspective, GW observations of binary systems alone are unable to distinguish between the scale-independent EMSG and GR, at least up to the PN order applied here. We refer readers to~\cite{2022arXiv221004668A} for further discussions on such features of the EMPG model.

Therefore, to utilize the interpretation of GW observations as a practical tool to test this modified theory, it is necessary to have additional information from physically independent phenomena, for instance, from some other astrophysical events that can provide us with additional measurements of the masses of the two components of binary systems with required accuracy, from cosmological data that can constrain the free parameter of the scale-independent EMSG regarding its consequences on the dynamics of the universe (see, e.g.,~\cite{2018PhRvD..98f3522A} for a cosmological constraint of this gravity model), etc. In order to find the individual masses, one could go beyond the quadrupole formula~\eqref{hijN2prho} and obtain higher PN corrections to the gravitational potential in EMPG. It means that the higher PN corrections to the multipole moments as well as to the EMPG equations of motion for binary systems should be derived. In this case, similar to GR, the orbital phase of the GW signals would explicitly depend on the symmetric mass ratio of the binary system in addition to its chirp mass. Optimistically, using a full matched-filter analysis as introduced in~\cite{1992PhRvD..46.5236F,1994PhRvD..49.2658C}, it is then possible to determine the source parameters such as $\mathcal{M}$ and $\nu$ along with an accurate limit for the free parameter of the scale-independent EMSG. It is worth mentioning that Ref.~\cite{1994PhRvD..50.6058W} applies the matched-filter analysis to constrain the parameter of Brans-Dicke theory. Such a full analysis is beyond the scope of the current paper and we leave it to future works in this context.

\subsection{Tests from binary pulsar observations}

In the sense of the strong-field gravity, another type of data that can provide us with a rich test of the modified theories of gravity is binary pulsar observations.  This kind of observation has its own merits. In fact, since several relativistic effects can be accurately measured in relativistic binary pulsar systems, one a priori expects that to place tight constraints on the free parameters of gravitational theories is possible.    
Therefore, it is worthwhile to test the scale-independent EMSG using binary-pulsar experiments as well.

Here, we focus our attention on the best-observed post-Keplerian parameter, i.e., the first time derivative of the orbital period of binary systems. In Subsec.~\ref{Radiative losses}, the scale-independent EMSG version of this parameter is introduced. Moreover, among several known relativistic binary pulsars, we choose the system whose orbital period change is most precisely measured. It is shown in~\cite{2021PhRvX..11d1050K} that the observed change in the orbital period of the double pulsar PSR J$0737-3039$A/B due to GW emission, $\dot{P}_{\text{obs}}$, is equal to
\begin{align}\label{P_obs}
\dot{P}_{\text{obs}}=-1.247782(79)\times 10^{-12},
\end{align}
and its ratio to the one predicted in GR, $\dot{P}_{\text{GR}}$, is given by \footnote{See equations (44) and (48) of~\cite{2021PhRvX..11d1050K}, respectively.} 
\begin{align}\label{precise_ratio}
\dot{P}_{\text{obs}}/\dot{P}_{\text{GR}}=0.999963(63).
\end{align}  
Here, numbers in parentheses are $1\sigma$ uncertainties in the last two digits. It is the most precise test of GW emission obtained so far from binary pulsars~\cite{2021PhRvX..11d1050K}. Compared to the Hulse-Taylor binary pulsar, this value is about an order of magnitude better, cf.~\cite{2016ApJ...829...55W}. It is worth mentioning that to obtain the result~\eqref{P_obs}, the effect of the Galactic and Shklovskii accelerations  as well as the mass loss contributions are included in the calculations.

This high precision allows us to place a tight constraint on the free parameter of the scale-independent EMSG. In order to find this limit, we utilize the same analysis applied by~\cite{2013MNRAS.431..741D,2022PhRvD.105d4014N}. It is assumed that the scale-independent EMSG modification of $\dot{P}$ can completely justify the observed change in the orbital period of the binary system due to GW emission. Doing so, we have $\dot{P}_{\text{\tiny EMPG}}=\dot{P}_{\text{obs}}$. Keeping this fact in mind and regarding Eq.~\eqref{Tdotp}, we get to 
\begin{equation}
\label{eq101}
\alpha=-\frac{3}{8}\Big(\frac{\dot{P}_{\text{obs}}}{\dot{P}_{\text{GR}}}-1\Big),
\end{equation}
for the EMPG free parameter for the case  with  $\mathcal{L}_{\text{m}}=-\varepsilon c^2$. Here, 
\begin{equation}
\begin{aligned}
\dot{P}_{\text{GR}}=&-\frac{192\pi}{5}\Big(\frac{G \mathcal{M}}{c^3}\frac{2\pi}{P}\Big)^{\frac{5}{3}}\\
&~~~~~~~~~~~~~~~~~~~~\times\big(1-e^2)^{-\frac{7}{2}}\Big[1+\frac{73}{24}e^2+\frac{37}{96}e^4\Big].
\end{aligned}
\end{equation}
Therefore, only if $\alpha$ satisfies the condition~\eqref{eq101}, the scale-independent EMSG can pass this strong-field gravity test with flying colors. Now, using this result and the constraint given in~\eqref{precise_ratio}, one can reach the following range for the free parameter of the theory under consideration:
\begin{align}\label{limit_p}
-1.0\times 10^{-5}<{\alpha}<3.8\times 10^{-5}.
\end{align}  
This bound is more constraining than what can be obtained from the GW events GW190425 and GW170817 mentioned in the preceding section.

As a final point, it should be emphasized that in order to derive this limit, we use the mass values obtained by assuming the validity of GR and utilizing the well-measured Keplerian parameters. However, in a standard procedure, one should investigate at least two post-Keplerian parameters in the gravity theory under consideration to calculate the two (a priori unknown) masses of binary systems, e.g., in the GR case, see~\cite{2004hpa..book.....L}. In other words, one can truly compare the observed $\dot{P}$ with the predicted one only when the chirp mass (or the mass of each component) is known under the assumption of the asked gravity theory.
Nevertheless, in the absence of a complete analysis of post-Keplerian parameters in the scale-independent EMSG giving us the component masses of binary pulsars, as a preliminary step towards constraining this theory in the strong-gravity regime, we apply the results in equations (36), (37), and (47) of~\cite{2021PhRvX..11d1050K}. It should be noted that the latter result includes 3.5\tiny PN \normalsize corrections to the equations of motion of the binary system. In the sequel of current work, we aim to determine the mass of the components in the binary pulsar system and perform a self-consistency test of the scale-independent EMSG by studying the relativistic properties of this theory at least in three suitable post-Keplerian parameters.

\section{Summary and conclusions}\label{Summary}

In this paper, the radiative properties of the scale-independent energy-momentum squared gravity (EMSG)~\cite{2018PhRvD..98f3522A}, which corresponds to a particular case $\eta=\frac{1}{2}$ of the energy-momentum powered gravity (EMPG)~\cite{2018PhRvD..97b4011A,2017PhRvD..96l3517B}, have been studied. To do so, similar to the previous work~\cite{2022PhRvD.105d4014N}, where the particular case $\eta=1$ of EMPG is investigated, we have utilized the Landau-Lifshitz formulation of the field equations. 
It has been assumed that the matter source is subjected to the slow-motion condition and the weak-field limit.
Depending on the choice of the matter Lagrangian densities, this type of modified theory of gravity could predict different results. To show this fact, we have considered the cases $\mathcal{L}_{\text{m}}=-\varepsilon c^2$ and $\mathcal{L}_{\text{m}}=p$, both of which describe a perfect fluid. In fact, given these two Lagrangian densities, we have encountered two different theories in the framework of EMPG. 

The field equations of the model have been solved approximately to the leading post-Newtonian (PN) order. Using the results~\eqref{hijN2prho}, the gravitational potentials and consequently the gravitational-wave (GW) field have been studied in the wave zone.  Considering a binary system as a source of GW signals, it has been revealed that, this theory with $\mathcal{L}_{\text{m}}=-\varepsilon c^2$ represents different predictions for GW emissions compared to GR ($\alpha=0$), while  in the case $\mathcal{L}_{\text{m}}=p$, the EMPG gravitational potential does not deviate from what is given in GR and higher PN corrections should be investigated in this regard. So, depending on the choice of $\mathcal{L}_{\text{m}}$, the results are different.  In the case $\mathcal{L}_{\text{m}}=-\varepsilon c^2$, up to the PN order considered here, it has been shown that the scale-independent EMSG only modifies the GW amplitude and leaves its scale-free polarizations unchanged. Furthermore, we have found that this modified gravitational radiation extracting the orbital energy of the binary system, leads to a secular change of the Keplerian parameter $P$ containing new terms due to the scale-independent EMSG. Therefore, the radiative property of this gravity theory would manifest itself in the corresponding post-Keplerian parameter/radiative parameter, i.e., the first time derivative of the orbital period of the binary system. These non-GR contributions to $\dot{P}$ have been presented in Eq.~\eqref{Tdotp}.

Next, we have implemented our results along with the GWs observations from the relativistic binary systems to constrain the dimensionless parameter $\alpha$, the only free parameter of the scale-independent EMSG quantifying its deviation from GR. This gives us a limit for the model with $\mathcal{L}_{\text{m}}=-\varepsilon c^2$. Two types of observations can be applied to do so:  GW observations of inspiralling binaries and binary pulsar observations called the direct and indirect detections of GWs, respectively. From the former, using a crude analysis, estimating the accumulated phase of GWs, we have found that from the GW event GW170817, the EMPG free parameter should be within the following bound:  $\arrowvert{\alpha}\arrowvert<4.5\times10^{-5}$.
On the other hand, from the latter and regarding the observed change in the orbital period of the double pulsar PSR J0737$-$3039A/B, it has been obtained that $-1\times 10^{-5}<{\alpha}<3.8\times 10^{-5}$. Comparing these results shows that the current measurement of this binary pulsar gives tighter constraints on $\alpha$. This limit is the most important result of this paper. This is the first study that constrains the free parameter of EMPG with $\mathcal{L}_{\text{m}}=-\varepsilon c^2$. We have also shown that to constrain the free parameter of the model with $\eta=1/2$ and $\mathcal{L}_{\text{m}}=p$, higher PN corrections should be taken into account.
Studying a cosmological scenario and using the CMB Planck data and baryonic acoustic oscillations data, Ref.~\cite{2018PhRvD..98f3522A} show that $\alpha\sim10^{-7}$ for the case $\mathcal{L}_{\text{m}}=p$. The fact that the constraints on $\alpha$ from phenomena related to completely different scales of energy density, time, and length are of similar order of magnitude offers observational confirmation that a hallmark of the scale-independent EMSG, its deviation from the GR, remains effective regardless of the energy density scale, in contrast to many modified gravity theories in the literature.

It should be emphasized that from the practical point of view, direct and indirect GW observations of binary systems alone are indeed unable to distinguish between the scale-independent EMSG and GR ($\alpha=0$), at least up to the PN order applied in this work. According to our calculations, the extra radiative effects due to the new terms arisen from the scale-independent EMSG can be absorbed in the definition of the chirp mass of the binary system, giving a rescaled chirp mass. So, the relevant equations in the scale-independent EMSG would become mathematically identical to the GR ones. On the other hand, since the chirp mass of the binary systems are measured purely via GW observations, one cannot measure this rescaling and consequently the radiative effects of the scale-independent EMSG is utterly undetectable. In fact, to the order of accuracy considered here, the scale-independent EMSG model escapes this strong-field gravity test. Given this point and also the results given in~\cite{2022arXiv221004668A}, one may deduce that the non-GR contributions of this theory to the relevant relations in the weak- and strong-field regimes are similar; and it seems that only the gravitational mass would be rescaled/modified. However, keeping in mind that the new terms arisen from the scale-independent EMSG can be deeply buried in the higher PN orders, which are significant in the strong-field regime, it is still possible to find strong-field deviations from GR. To do so, in the case of direct GW observations from inspiralling binaries, a full matched-filter analysis can be performed to dig deeply into higher PN corrections. So, it may allow one to determine the scale-independent EMSG and GR versions of chirp mass, to constrain accurately the free parameter $\alpha$, and finally to test this modified theory of gravity in the strong-field regime. Furthermore, knowing that several relativistic and radiative effects can be accurately measured in these types of systems, a well-measured binary pulsar can be chosen as a suitable candidate to obtain the scale-independent EMSG and GR chirp masses and to distinguish between these two gravity theory in the strong-field regime without the need for extra information from other cosmological and astronomical observations or phenomena. To do this, a complete analysis of post-Keplerian parameters can be performed in the framework of the scale-independent EMSG.

To sum up, we have tested the scale-independent EMSG and constrained its free parameter $\alpha$ in the strong-field regime. After assuming that GR is the valid gravity theory, as a priori expectation, an interval has been obtained for $\alpha$. We should keep in mind that to find a more accurate estimate of $\alpha$, it is necessary to measure the mass of the two components of binary systems in this theory, either by using a full matched-filter analysis in the case of direct GW observations or by studying at least two suitable post-Keplerian parameters in the case of indirect GW observations. Of course, the validity of this limit should also be checked with other experiments. We leave this kind of thorough study for future works.

\section*{acknowledgments}
Helpful comments by the anonymous referee are gratefully acknowledged.
The authors thank Nihan Kat{\i}rc{\i} and N. Merve Uzun for useful discussions. \"{O}.A. acknowledges the support by the Turkish Academy of Sciences in the scheme of the Outstanding Young Scientist Award  (T\"{U}BA-GEB\.{I}P), and the COST Action CA21136 (CosmoVerse). E.N. and M.R. acknowledge the support by Ferdowsi University of Mashhad.  E.N. would like to thank
Shahram Abbassi for his continuous encouragement and support during this work.

\appendix

\section{Field equation solutions}\label{app1}

In this appendix, for the sake of convenience, we summarize the structure of $h_{\mathcal{N}}^{\alpha\beta}$ and $h_{\mathcal{W}}^{\alpha\beta}$ which are the solutions of the wave equation~\eqref{fieldeq_LL}. For detailed calculations to find these solutions, we refer the reader to~\cite{poisson2014gravity}.

The structure of the near-zone portion $h_{\mathcal{N}}^{\alpha\beta}$ is given by 
\begin{align}
 \nonumber
 h_{\mathcal{N}}^{\alpha\beta}{(t,\boldsymbol{x})}=&\frac{4G}{c^4}\sum_{l=0}^{\infty}\frac{(-1)^l}{l! c^l}\\\label{hNear}
&\times\Big(\frac{\partial}{\partial t}\Big)^l
\int_{\mathfrak{M}}\tau^{\alpha\beta}_{\text{eff}}{(t,\boldsymbol{x}')}\rvert{\boldsymbol{x}-\boldsymbol{x}'}\rvert^{l-1} {\rm d}^3x',
\end{align} 
for the case $x<\mathcal{R}$ and $x'<\mathcal{R}$.
Here, $\mathfrak{M}$ represents a three-dimensional sphere with radius $\mathcal{R}$ separating the near and wave zones.   
For the next case, we have
 \begin{align}\label{hNwave}
h_{\mathcal{N}}^{\alpha\beta}(t,\bm{x})=&\frac{4G}{c^4}\sum_{l=0}^{\infty}\frac{(-1)^{l}}{l!}
 \\\nonumber
 &\times\partial_{j_1j_2\cdots j_l}\bigg[\frac{1}{r}\int_{\mathfrak{M}}\tau^{\alpha\beta}_{\text{eff}}(\tau,\bm{x}')x'^{j_1j_2\cdots j_l}{\rm d}^3x'\bigg],
 \end{align}
where the field point is located in the wave zone, i.e., $x>\mathcal{R}$,  
$x^{j_1j_2\cdots j_l}$ stands for $x^{j_1}x^{j_2}\cdots x^{j_l}$, and $\partial_{j_1j_2\cdots j_l}$ shows $\partial_{j_1}\partial_{j_2}\cdots \partial_{j_l}$. Here, $r=|\bm{x}|$ and $\tau=t-r/c$ is the retarded time. The integrand in this integral unlike the previous solution is a function of $\tau$.
 
The structure of $h_{\mathcal{W}}^{\alpha\beta}$ that we need during our calculation, is equal to 
 \begin{align}\label{eq8}
 \nonumber
  h^{\alpha\beta}_{\mathcal{W}}(t,\bm{x})&=\frac{4 G}{c^4}\frac{n^{\langle j_1j_2\cdots j_l \rangle}}{r}\bigg\lbrace\int_{0}^{\mathcal{R}}f^{\alpha\beta}(\tau-2s/c)A(s,r){\rm d}s\\
 &+\int_{\mathcal{R}}^{\infty}f^{\alpha\beta}(\tau-2s/c)B(s,r){\rm d}s\bigg\rbrace,
 \end{align}
where the field and source points are situated in the wave zone. Here,  $A(s,r)=\int_{\mathcal{R}}^{r+s}P_{l}(\zeta)p^{1-n}{\rm d}p$ and $B(s,r)=\int_{s}^{r+s}P_{l}(\zeta)p^{1-n}{\rm d}p$ in which $P_{l}(\zeta)$ is a Legendre polynomial and $\zeta=(r+2s)/r-2s(r+s)/(rp)$.
To obtain Eq. \eqref{eq8}, the source function is restricted to the form   
\begin{align}\label{eq7}
\tau^{\alpha\beta}=\frac{1}{4\pi}\frac{f^{\alpha\beta}(\tau)}{r^n}n^{\langle j_1j_2\cdots j_l \rangle}.
\end{align}
In the above relations, $n^{\langle j_1j_2\cdots j_l \rangle}$ is an angular STF tensor. See equation (1.154) of~\cite{poisson2014gravity} for its definition.
It should be noted that at higher PM order, $\tau^{\alpha\beta}$ has logarithmic forms, and consequently Eq. \eqref{eq8} has to be generalized. As we focus on the leading order in the present work, the above structure of $h^{\alpha\beta}_{\mathcal{W}}$ is sufficient.

\end{document}